 \documentclass[final,3p,times,twocolumn]{elsarticle}

\usepackage{amssymb}
\usepackage[table]{xcolor}
\usepackage[colorlinks]{hyperref}
\usepackage{soul}

\usepackage{amsmath}
\DeclareMathOperator{\erf}{erf}
\usepackage{booktabs}
\usepackage{multirow}
\usepackage{gensymb}

\journal{Journal of Wind Engineering \& Industrial Aerodynamics }

\newcommand{\hide}[1]{}

\begin{document}
\begin{frontmatter}


\title{Kirigami-inspired wind steering for natural ventilation}

\author[addr]{Lucia Stein-Montalvo}
\author[addr2]{Liuyang Ding}
\author[addr2]{Marcus Hultmark}
\author[addr]{Sigrid Adriaenssens}
\author[addr]{Elie Bou-Zeid\corref{mycorrespondingauthor}}
\cortext[mycorrespondingauthor]{Corresponding author}
\ead{ebouzeid@princeton.edu}

\address[addr]{Department of Civil and Environmental Engineering, Princeton University, Princeton, NJ 08540, USA}
\address[addr2]{Department of Mechanical Engineering, Princeton University, Princeton, NJ 08540, USA}

\begin{abstract}
Ensuring adequate ventilation of exterior and interior urban spaces is essential for the safety and comfort of inhabitants. Here, we examine how angled features can steer wind into areas with stagnant air, promoting natural ventilation. Using Large Eddy Simulations (LES) and wind tunnel experiments with particle image velocimetry (PIV) measurements, we first examine how louvers, located at the top of a box enclosed on four sides, can improve ventilation in the presence of incoming wind. By varying louver scale, geometry, and angle, we identify a geometric regime wherein louvers capture free-stream air to create sweeping interior flow structures, increasing the Air Exchange Rate (ACH) significantly above that for an equivalent box with an open top. We then show that non-homogeneous louver orientations enhance ventilation, accommodating winds from opposing directions, and address the generalization to taller structures. Finally, we demonstrate the feasibility of replacing louvers with lattice-cut \textit{kirigami} (``cut paper"), which forms angled chutes when stretched in one direction, and could provide a mechanically preferable solution for adaptive ventilation. Our findings for this idealized system may inform the design of retrofits for urban structures -- e.g. canopies above street canyons, and ``streeteries” or parklets -- capable of promoting ventilation, while simultaneously providing shade.

\end{abstract}

\end{frontmatter}


\section{Introduction}\label{sect:intro}
The rapid pace of urbanization is closely intertwined with several pressing issues, including elevated pollutant levels, the threat of future global pandemics, and rising global temperatures \citep{IPCC6-urb}. Thus, it is imperative for cities to address and adapt to these simultaneous challenges \citep{Gonzalez21}.

At the regional to city-scale, urban areas are prone to sub-optimal ventilation patterns due to strong drag by buildings, with implications such as the entrapment of wildfire smoke and urban heat~\cite{Britter2003,LlagunoMunitxa2018,Ng2011,Xu2017}. Within cities, urban street canyons can collect and recirculate contaminants, putting inhabitants at risk of adverse health effects~\cite{Oke1988}. As a result, canyon ventilation has received significant research attention, with studies considering the effects of factors including canyon aspect ratio~\cite{Oke1988,Liu2005,Nazridoust2006,Li2019},  height asymmetry~\cite{Li2020,Klukov2021}, thermal effects~\cite{Kim1999,Nazarian2016}, and environmental factors like atmospheric turbulence~\cite{Salizzoni2009}, wind speed~\cite{Nazridoust2006}, and wind direction~\cite{Soulhac2007}. Building-scale urban morphology has been shown to play a dominant role in canyon ventilation, leading to the investigation of geometric features like roof shape~\cite{Huang2014,Wen2018} -- which can be tuned to promote ventilation of urban streets~\cite{LlagunoMunitxa2017,LlagunoMunitxa2018} -- as well as building façades and balconies~\cite{LlagunoMunitxa2018,Zheng2022}, building roughness~\cite{Fellini2020}, and obstacles within the street~\cite{Buccolieri2022}, including vegetation~\cite{Giometto2017,Li2018,Huang2019}.

\begin{figure*}[t] 
	\centering
	\includegraphics[width=0.8\linewidth]{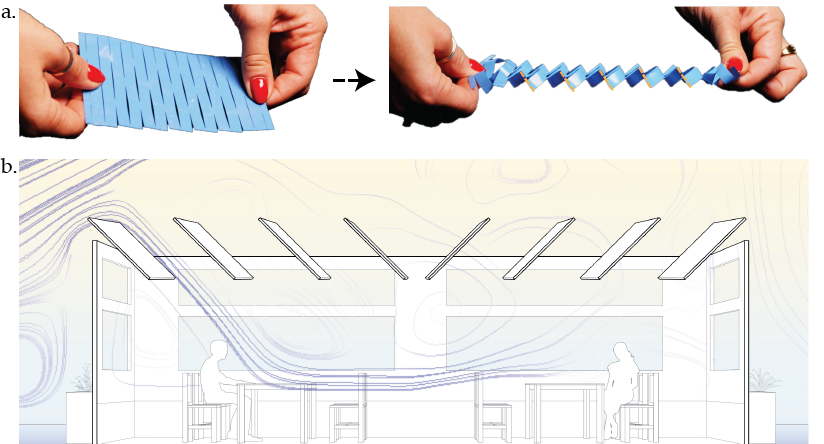}
	\caption{Illustration of the concept for kirigami-inspired wind-steering. a. As a sheet patterned with lattice cuts is stretched, unit cells buckle out of plane and rotate. Angled chutes then can redirect oncoming wind. b. Louvers, as a quasi-2D approximation of the complex kirigami geometry, could adaptively provide shade and promote ventilation in urban canyon settings, including a ``streetery" or parklet, as shown conceptually.
	} 
	\label{fig:motivation}
\end{figure*} 

A wealth of research concerning natural, or passive, ventilation of interior spaces has also emerged over the past few decades~\cite{Ohba2010,Etheridge2015}, motivated primarily by the desire to improve air quality and thermal comfort, while maintaining low energy consumption. Further, a recent surge of research on the topic has been fueled by public health concerns regarding the transmission of airborne viruses~\cite{Morawska2020,Tang2020,Bhagat2020}. Natural ventilation is driven by pressure or temperature gradients, and thus is sensitive to the locations and sizes (i.e. porosity) of building openings~\cite{Seifert2006,Karava2007,Tominaga2016,Izadyar2020}, thermal gradients~\cite{Linden1999,Khanal2011,DaviesWykes2020}, and outdoor microclimate conditions~\cite{Ai2015} including wind speed and direction~\cite{Chu2023}, which are sensitive to surrounding buildings (i.e. sheltering conditions)~\cite{Shirzadi2019}. Chimney-like structures aimed at enhancing natural ventilation and cooling by drawing wind into buildings, known as ``wind catchers", have been used for centuries in Persian Gulf countries~\cite{Jomehzadeh2017,Saadatian2012}. The potential for modern, commercial use of similar ``wind catching'' concepts has been studied in recent years~\cite{Hughes2012}, and similar designs -- i.e. structures protruding above the canyon to direct wind downward -- have even been suggested for improving outdoor canyon ventilation~\cite{Chew2017,Zhang2019,Ming2020,Lauriks2021}.

Meanwhile, as the global climate warms -- and more so in cities~\cite{Li2013,Manoli2019} -- engineered shading solutions, e.g. canopies above the urban canyon, are increasingly recognized as another way to improve thermal comfort at low energy cost~\cite{Middel2021}. Furthermore, as many people and business moved activities ``outside" during the Covid-19 pandemic, structures designed to provide shade and/or shelter, i.e. ``streeteries" or parklets (Fig.~\ref{fig:motivation}b) emerged en masse in cities across the United States and other countries. However, many of these enclosed designs could likely further impede indoor-outdoor air exchange, negating some of the sought-after health benefits. Thus, cities need new retrofitting designs that can both provide shade and improve ventilation and that can be applied to interior (buildings) and exterior (urban) spaces.

\begin{figure*}[t] 
	\centering
	\includegraphics[width=\linewidth]{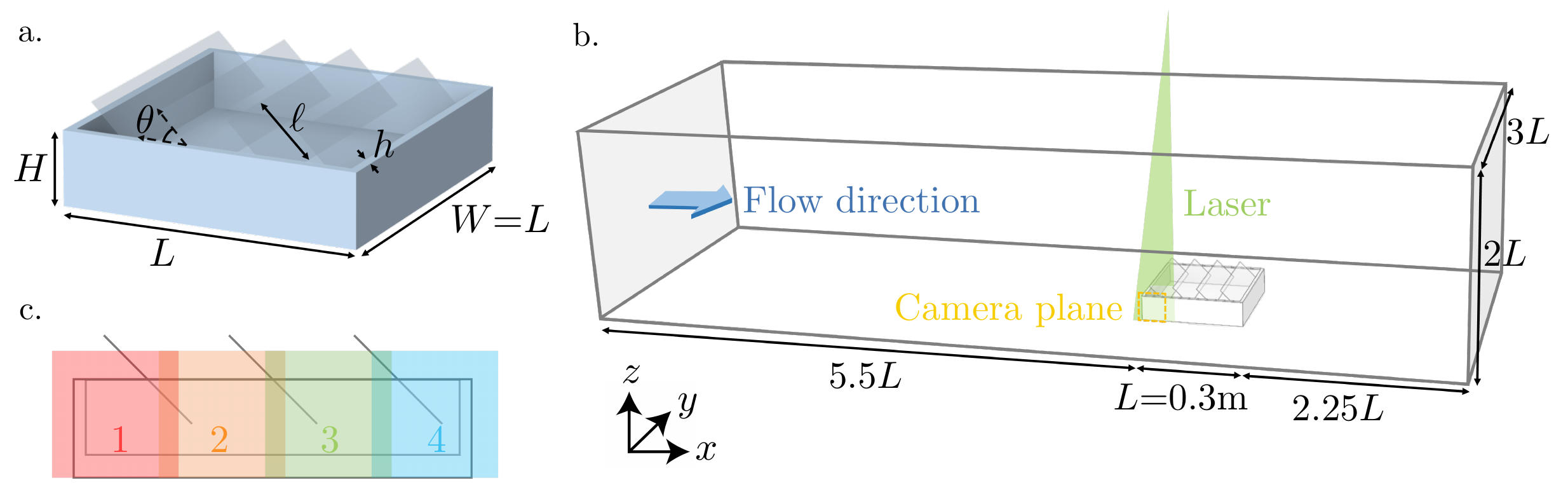}
	\caption{Schematic of the geometric parameters and experimental setup. a. Geometric parameters of the box with $N=3$ louvers. In experiments and simulations, $L = W \approx 30.5$ cm, $h \approx 9.5$ mm (the inner wall length $L_{\text{in}} = L - 2h$), and $\ell = L_{\text{in}}/N$ (here, $N=3$, and $\ell \approx 9.5$ cm). The louver angle $\theta = 45 ^\circ$ is shown. The box height $H \approx 7.5$ cm in experiments and $H \approx {7.5,13.7,26.4}$ cm in simulations.
		b. Schematic of the test section of the closed-loop wind tunnel. Seeded air circulates, illuminated by the laser sheet. The camera captures 2000 images of the region indicated by the yellow dashed square (location 1 inside the box) at a rate of 25 Hz (80 s). The box then is shifted in increments of $80$ mm to three other positions (locations 2-4) to construct the full plane of view inside the box. The coordinate system is shown; the origin is located at base of the domain, in the $x-y$ center of the box.
		c. Indication of the four view locations along the box. 
	} 
	\label{fig:geom}
\end{figure*} 

In this work, we investigate how angled features, which could reasonably be expected to provide shade, can steer wind into spaces with stagnant air to improve ventilation. 
We use Large Eddy Simulations (LES) and wind tunnel experiments with particle image velocimetry (PIV) measurements to examine the ventilation patterns of a 
simple, idealized system: a standalone, shallow box with a louvered ``roof", in the presence of incoming wind. While this system resembles a model building, enclosed streetery, or interior courtyard, our aim is to understand fundamental wind steering mechanisms that could be applied to retrofitting designs for interior or exterior spaces, given a proper characterization of air flow patterns in a specific urban setting. In Sect.~\ref{sect:geommethods}, we elaborate on our methods and the range of parameters studied, before validating our simulations against experiments in Sect. ~\ref{sect:validate}. We present the results of a parametric study in which we vary the louver size and angle in Sect.~\ref{sect:results}, where we study ventilation through the Air Exchange Rate (ACH) and an efficiency parameter based on an ideal wind capture scenario. In Sect.~\ref{sect:optimize}, we show how incorporating louvers with nonhomogenous orientation (as shown schematically in Fig.~\ref{fig:motivation}b) can further improve ventilation, and accommodate wind from different directions. We then briefly discuss the implications of taller box geometries, which are likely more relevant to urban canyon ventilation~\cite{Hunter1990}, in Sect.~\ref{sect:height}. 

Guided by these results, we then turn to investigate an alternative solution for adaptive ventilation based on \textit{kirigami}, or “cut-paper." In kirigami, adding cuts to thin sheets frees sections to buckle and tilt when stretched, transforming a stiff, 2D structure into a flexible, 3D one (see Fig. ~\ref{fig:motivation}a). Recently, this ancient Japanese art form has earned popularity in engineering research. Simply by modifying the cut patterns in a sheet, engineers can control the mechanical behavior and motions exhibited~\cite{Callens2018,Holmes2019,Tao2022} across scales~\cite{Dias2017}, enabling the design of functional structures ranging from robotics devices~\cite{Rafsanjani2018, Yang2021} to adaptive shading~\cite{Tang2016,Yi2018,Arauz2023} façades with advanced thermal regulation properties~\cite{Ke2019,Yin2023}. We propose that in addition to allowing airflow through openings, as was recently suggested~\cite{Arauz2023}, the angled pores that emerge when lattice-cut kirigami is stretched can also be designed to controllably redirect air flow. Furthermore, kirigami offers several potential benefits over louvers for practical use, e.g. single degree-of-freedom actuation via uniaxial stretching, and a readily tunable force response by changing the cut patterns and amount of stretch. Thus, in Sect.~\ref{sect:kirigami}, we use PIV and LES to demonstrate the feasibility of using kirigami for ventilation, before offering our concluding remarks in Sect.~\ref{sect:conclusions}. This work will serve as a proof-of-concept, and offer a pathway for future optimization of kirigami for wind steering. Furthermore, though care is necessary to differentiate the open box geometry from realistic urban settings, we expect that our findings for this idealized system can inform the design of structures aimed at improving natural ventilation in realistic urban settings, including adaptable shading canopies above the urban canyon, pavilions, and streeteries.

\section{Geometry and methods}\label{sect:geommethods}
To learn how louvers interact with wind in experiments and simulations, we study an idealized system consisting of a shallow, isolated box enclosed on four walls but open on the top (see Fig.~\ref{fig:geom}) immersed in an incoming wind. The box used for experiments and baseline simulations has external dimensions $L=W=30.48$ cm $\times$ $H=7.47$ cm. (The height was varied in some simulations discussed later.) The box dimensions were selected to be as large as possible while ensuring blockage was sufficiently small in the wind tunnel; the cross-section blockage percentage is $4.1 \%$. The walls and the base are $h = 9.525$ mm thick, so that the internal volume is $V = 5.322$ mm$^3$. $N \geq 0$ louvers of width $\ell$ are placed with an axis of rotation at the top plane of the box (HP2, see Fig.~\ref{fig:les}), and their size is constrained so that when the louvers are flat, i.e. $\theta = 0$, they fully cover the top of the box, i.e. $\ell = L_{in}/N$ where $L_{in} = L-2h$. Additional simulations were then performed with $H = 0.14$ m and $H = 0.26$ m for $N=3$ and $N=0$, as discussed in Sect.~\ref{sect:height}. The open box, i.e. $N=0$, serves as our reference case, representing a space with no shading or wind-steering structures. 

The geometry of the wind tunnel test section is reproduced in simulations. We define a coordinate system as shown in Fig.~\ref{fig:geom}c: $x$ is the horizontal streamwise direction, $y$ is horizontal and transverse to the main flow direction, and $z$ points upward in the vertical direction. The origin is at the center of the box in the $x-y$ plane, and at the base of the domain (i.e. where the base of the box meets the bottom surface of the wind tunnel). 
The louver parameters studied are summarized in Table 1, including the reference mean streamwise inflow velocity $\overline{u}_{\text{in}}$, which is measured at a distance $0.05 L$ upstream (to the left in all plots) and $0.1 H$ above the box, at $y=0$ (i.e. the middle of the box, width-wise). We note that the velocity $\overline{u}_{\text{in}}$ differs from the further-upstream inlet velocity $u_{\infty}$, which is set to 3.5 m/s in simulations and is not measured in experiments. Further details about experiments and simulations are provided next.

\begin{table}[h!]
	\label{table:parameters}
	\centering
	\begin{tabular}{ccccc}
		\toprule
		\multicolumn{1}{c}{$N$} & \multicolumn{1}{c}{$\ell$ (cm)} & \multicolumn{1}{c}{$\theta$ (degree)}& \multicolumn{1}{c}{$H$ (cm)} & $u_{in}$ (m/s)  \\
		\midrule
		\rowcolor[gray]{0.93}
0 & -- & -- & 7.47 & $5.50$ \\
\rowcolor[gray]{0.93}
3 & 9.5 & 45 (all) & 7.47 & $3.26$ \\
\rowcolor[gray]{0.93}
3 & 9.5 & -45 (all) & 7.47 & $3.80$ \\
\rowcolor[gray]{0.93}
3 & 9.5 & 45, 45, -45 & 7.47 & $2.35$ \\
\rowcolor[gray]{0.93}
3 & 9.5 & 45, -45, -45 & 7.47 & $2.83$ \\
\rowcolor[gray]{0.8}
		0 & --  & --& 7.47 & $2.50$ \\
		\rowcolor[gray]{0.8}
		0 & --  & --& 13.65 & $2.83$ \\
		\rowcolor[gray]{0.8}
		0 & --  & --& 26.35 & $2.51$ \\
		\rowcolor[gray]{0.8}
		2 & 14.29 & 45 (all) & 7.47 & $2.28$ \\
		\rowcolor[gray]{0.8}
		3 & 9.53 & 22.5 (all) & 7.47 & $2.50$ \\
		\rowcolor[gray]{0.8}
		3 & 9.53& 45 (all)& 7.47 & $2.59$ \\
		\rowcolor[gray]{0.8}
		3 & 9.53& 45 (all)& 13.65 & $2.60$ \\
		\rowcolor[gray]{0.8}
		3 & 9.53& 45 (all)& 26.35 & $2.40$ \\
		\rowcolor[gray]{0.8}
		3 & 9.53& 67.5 (all)& 7.47 & $2.35$ \\
				\rowcolor[gray]{0.8}
		3 & 9.53& 90 (all)& 7.47 &  $2.41$ \\
		\rowcolor[gray]{0.8}
			4 & 7.14 & 45 (all)& 7.47 & $2.52$ \\
			\rowcolor[gray]{0.8}
					5 & 5.72 & 45 (all)& 7.47 & $2.58$ \\
					\rowcolor[gray]{0.8}
					6& 4.76 & 45 (all)& 7.47 &$2.61$ \\
					\rowcolor[gray]{0.8}
					7& 4.08 & 45 (all )& 7.47 & $2.51$ \\
					\rowcolor[gray]{0.8}
					14& 2.04 & 45 (all )& 7.47 & $2.43$ \\
						\bottomrule
	\end{tabular}
\caption{Parameter configurations studied in louver experiments (lighter gray rows) and simulations (darker gray rows). The columns refer to the number of louvers $N$, the louver width $\ell$, the louver angle $\theta$ (``(all)" refers to configurations with a homogeneous angle $\theta$ for all louvers), the box height $H$, and the mean streamwise inflow velocity $\overline{u}_{\text{in}}$, measured at $0.05L$ upstream and $0.01H$ above the box, at $y=0$. 
}
\end{table}

\subsection{Wind tunnel PIV experiments}\label{sect:wtmethods}
\begin{figure}[h!] 
	\centering
	\includegraphics[width=0.75\linewidth]{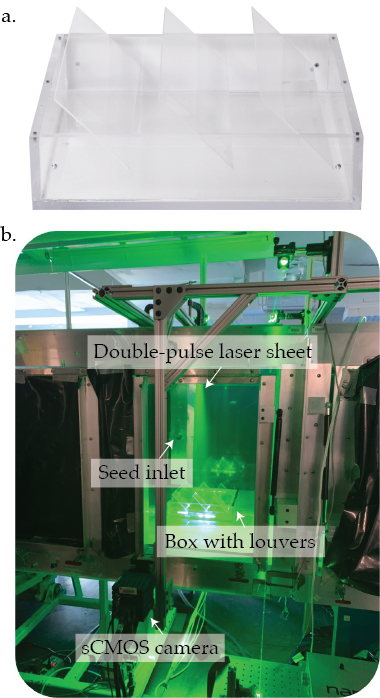}
	\caption{Photos of the experimental setup. a. Acrylic box with $N=3$ louvers used in experiments. b. Experimental setup for taking PIV measurements in the wind tunnel.
	} 
	\label{fig:expphotos}
\end{figure}

\begin{figure*}[t!] 
	\centering
	\includegraphics[width=\linewidth]{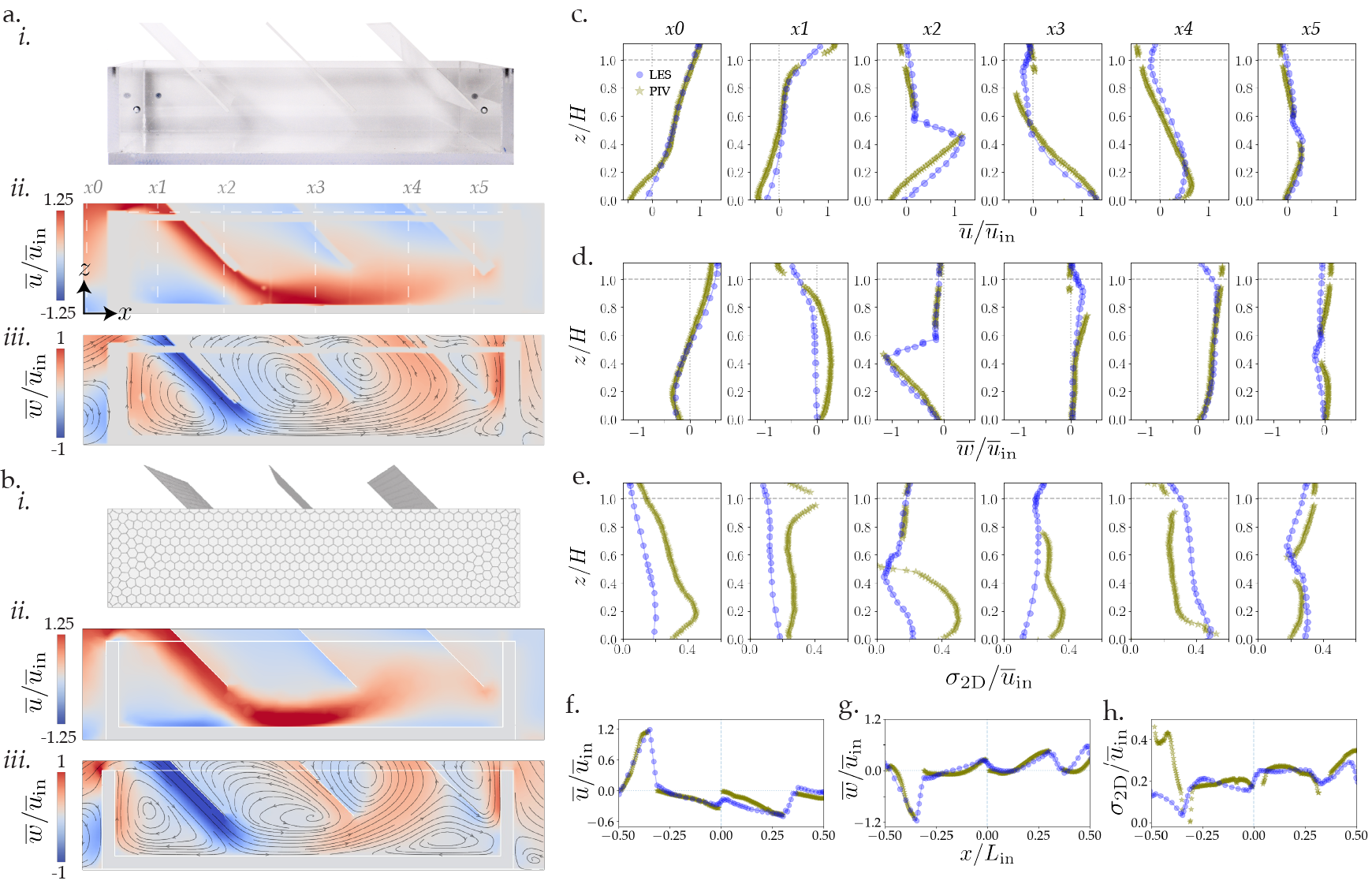}
	\caption{Validation of LES compared to PIV data from experiments for $N=3$ louvers (inflow from left.)
    a. Experimental setup (i.) and PIV results (ii.-iii.) 
    b. LES setup (i.) and results (ii.-iii.). The time-averaged streamwise velocity at VP1 is shown in the pseudocolor plots in a. ii. (PIV) \& b. ii. (LES), and the time-averaged vertical velocity is shown in a. iii. (PIV) and b. iii. (LES). 2D streamlines are shown in a.,iii. \& b.,iii. Vertical profile lines $x0-x5$, and the horizontal profile line, are delineated in a.,ii.
    c.-h. Profiles for $N=3$ louvers, comparing LES (blue dots) to PIV (green stars). Vertical profiles in c.-e. are taken at $x0 = -0.15$, $x1 = -0.12$, $x2 = -0.07$, $x3 = -0$, $x4 = 0.07$, and $x5 = 0.14$ m, as indicated with dashed gray lines in a.,ii. Horizontal profiles in f.-h. are at $z=H$, i.e. at the top of the box, as indicated with the dashed gray line in a., iii.
    c.\& f. Mean streamwise velocity. 
    d.\& g. Mean vertical velocity. 
    e.\& h. Total 2D ($x-z$) RMS velocity (calculated with respect to the time-averaged velocity.)
	} 
	\label{fig:validate}
\end{figure*}

The box used in experiments, shown in Figs.~\ref{fig:expphotos} and ~\ref{fig:validate}, was constructed from acrylic. Additional, thinner acrylic (thickness $1.6$ mm) sheets were laser cut with angled slots for mounting the louvers, and were interchangeably secured against to the inside of opposing side walls. Louvers were also laser cut from acrylic of the same thickness. For the kirigami experiment in Sect.~\ref{sect:kirigami}, a $1$ mm thick sheet of clear polyethylene terephthalate glycol (PETG) was laser cut with a pattern of linear, parallel cuts. The sheet, with length $25.4$ cm and width $30$ cm, was stretched to fit the same acrylic box used for louvers, and fixed with custom aluminum clamps. Clamps were secured by 3 mm screws, which were fed through small holes in the clamps and kirigami sheet, and into internally threaded holes in the wall thickness on opposite sides of the box. 

Wind tunnel experiments were performed in Princeton University's Instructional Fluid Dynamics Laboratory, which contains a modified version of a wind tunnel model 407-A manufactured by Engineering Laboratory Design, Inc. The test section has dimensions $0.61$ m (height) $\times$ $0.91$ m (width) $\times$ $2.64$ m (length), and transparent plexiglass walls. The flow passes through a conditioning area containing a honeycomb, a screen, and a rectangular 4.2:1 contraction area. An axial flow fan with variable pitch blades produce velocities in the range of $3-55$ m/s, and a heat exchanger upstream of the flow conditioning sections controls the air temperature~\cite{Jimenez2007}. 

\begin{figure}[h!] 
	\centering
	\includegraphics[width=\linewidth]{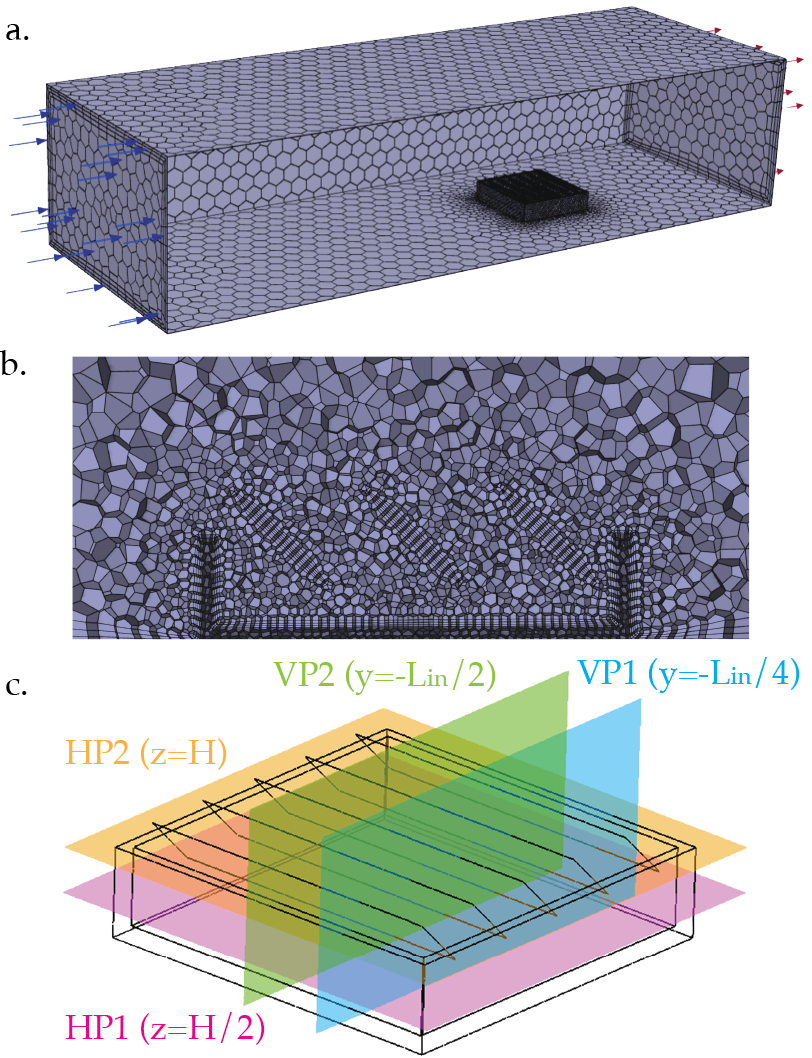}
	\caption{Simulation setup. a. Computational domain surface mesh, with refinement in the vicinity of the box. b. Side view of the mesh arrangement around the box with $N=3$ louvers at the vertical midplane (i.e. VP2, as shown in c.)
 c. Indication of data planes used for computation. 
	} 
	\label{fig:les}
\end{figure}

For experiments with louvers, we used PIV to characterize the flow field in an $x-z$ plane located at a quarter of the box width from the side (i.e., VP1, see Fig. \ref{fig:les}). Due to the nonhomogeneous geometry of kirigami, we performed an additional experiment at the vertical midplane of the box, i.e. VP3, by shifting the box in the y-direction. The flow was seeded with oil droplets (1 $\mu$m in mean diameter) generated by a Laskin nozzle type seeder (LaVision) and illuminated with a 1-mm thick light sheet from the top of the wind tunnel. The light sheet was formed by passing a laser beam generated by a dual-cavity Nd:YAG laser (Litron Nano 50 mJ) through sheet-forming optics (LaVision). The width of the light sheet at the height of the box was approximately 150 mm, and therefore particle images were acquired in four views with a streamwise separation of 80 mm between consecutive ones to cover the entire length of the box along VP1 (see Fig. \ref{fig:geom}c). The four views were achieved by successively shifting the box upstream so that the camera and laser optics could stay immobile, and each field view was 115 mm ($x$) $\times$ 95 mm ($z$). The location of the front of the box ranged from $5.5L$ to $4.7L$ from the end of the contraction, and PIV measurements were conducted in empty wind tunnel to characterize the inflow velocity profile at location 1, revealing a turbulent boundary layer depth of about $0.7H$ (see Appendix B.) 2,000 pairs of particle images were acquired at 25 Hz in each view using a LaVision Imager sCMOS camera (5.5 megapixels). Images were processed in DaVis 8.3.1 with an iterative multi-grid image deformation algorithm \citep{scarano2001iterative}. The final interrogation spot was 32 $\times$ 32 pixels with 50\% overlap, so that each velocity vector represents the in-plane ($x-z$ plane) velocity averaged over a 1.4 mm ($x$) $\times$ 1.4 mm ($z$) $\times$ 1 mm ($y$) volume, with a 0.7 mm spacing between vectors. 
 

Wind tunnel speed was controlled in experiments by setting the fan pitch to $0$, which corresponds to a free stream velocity of $U \approx 5$ m/s, as measured from PIV data (see Appendix B for typical velocity profiles.) This corresponds to a Reynolds numbers ($Re$) in the range of about $1 \times 10^4$ to $4 \times 10^4$, based on the length scale $H=0.0747$ m and an air viscosity of $1.5 \times 10^{-5}$ m$^2$ s$^{-1}$. Much higher velocities could not be sustained without damage to the acrylic louvers. We note that these $Re$ values are approximately one order of magnitude below what would be expected for a typical streetery of $2$ m height in $1$ m/s wind at pedestrian level (i.e. $Re_{streetery} \approx 1.3 \times 10^{5}$), and about two orders of magnitude below that for the top of a typical mid-level urban canyon of $10$ m in $2$ m/s wind. However, previous research suggests that the $Re$ in our experiments is sufficiently high so that the main flow features we focus on will not be affected much at larger $Re$ \citep{Khan2018,Khan2021}.


\subsection{Large Eddy Simulations}\label{sect:lesmethods}
To complement our louver experiments, obtain 3D fields for different parameter values, and easily vary geometry, we use Large Eddy Simulations (LES) to numerically simulate the flow and air exchanges. Compared to computationally-demanding direct numerical simulations (DNS), and Reynolds-averaged Navier Stokes (RANS) simulations that sacrifice accuracy and temporal resolution, LES offers efficient, high-fidelity turbulence modeling by filtering the governing Navier-Stokes equations so that all but the smallest turbulent flow structures are resolved. Thus, LES has become a widely used approach for many engineering applications, including for simulating urban flows~\cite{Li2016,Omidvar2020} and computing air exchange~\cite{LlagunoMunitxa2018}. We use the LES solver of Ansys Fluent 2021 R2 (Ansys Inc). For incompressible flow, the filtered continuity and momentum equations solved by Ansys Fluent LES reduce to:

\begin{equation}\label{eqn:les}
\begin{aligned}
\frac{\partial \tilde{u}_i}{\partial x_i} = 0, \\
\frac{\partial \tilde{u}_i}{\partial t} + \frac{\partial}{\partial x_j} (\tilde{u}_i \tilde{u}_j) = -\frac{1}{\rho} \frac{\partial \tilde{p}}{\partial x_i} + 2 \nu \frac{\partial \tilde{S}_{ij}}{\partial x_j} - \frac{\partial \tau_{ij}}{\partial x_j}.
\end{aligned}
\end{equation}
Here, the tilde ( $\tilde{}$ ) denotes filtering; $u_i$ is the velocity vector; $x_i$ is the position vector; $\rho$ is the fluid density; $\nu$ is the kinematic viscosity; $S_{ij}$ is the strain rate tensor; and $\tau_{ij}$ is the subgrid scale (SGS) stress tensor. Fluent models only the deviatoric part of $\tau_{ij}$ using an eddy viscosity ($v_t$) closure, i.e. $\tau_{ij}^D = -2 \nu_t \tilde{S}_{ij}$, and adds the isotropic part to the pressure. 

The geometry of the computational domain is defined to match that of the wind tunnel when the box is in location 1, and centered in the width direction. Thus, the enclosure dimensions are $0.61$ m (height) $\times$ $0.91$ m (width) $\times$ $2.64$ m (length), with the upstream wall of the box located $1.65$ m from the inlet. The box dimensions in simulations also match experiments, i.e. $L=W=0.3048$ m $\times$ $H=0.73$ m, with $0.95$ cm thick walls and base. $N = L_{\text{in}}/\ell$ louvers are created as baffle surfaces (no thickness), with their centerline at the top of the box, i.e. $y=H$. In simulations, there is a gap of $0.01$ m between the wall and louver edge. The louver angle $\theta$ is varied in $\{-45,22.5,45,67.5,90\} ^\circ$.

Fluent meshing is used to generate a polyhedral mesh, which is refined in the vicinity of the louvers and the box. The maximum element face area away from the box is $6.5 \times 10^{-3}$ m$^2$. Following a convergence study (see Appendix B), the local face size near the louver and box surface mesh was set to $7 \times 10^{-3}$ m (yielding a face area of $1.27 \times 10^{-7}$ m$^2$) for each geometry in our parametric study, with a growth rate of $1.2$. This results in $4$ to $24$ surface elements along the louver width depending on $\ell$ (i.e. each louver has fewer elements for smaller $\ell$, and vice-versa), $12$ on the box height, and between $7.7 \times 10^4$ and $9.8 \times 10^4$ volumetric elements in total. To resolve the viscosity-dominant regions near solid surfaces, three layers of small-thickness elements are added as a boundary layer in Fluent via the Add Boundary Layers task. As shown in the Appendix B, resolving the boundary layer improves agreement with experiments.

A laminar velocity boundary condition (i.e. zero turbulence intensity) was imposed at the inlet, and the inlet velocity was fixed at $u_{\infty} = 3.5$ m/s. Flow at the inlet is uniform, and a boundary layer develops over the length of the domain (see Appendix B.) In the convergence study discussed in Appendix B, we found no significant improvement in results compared to experiments when turbulent inflow was imposed instead. This is likely because the effect of the inflow conditions relaxes, while turbulence is generated at the wall and diffuses upward over the length of the domain. The inlet velocity corresponds to a Reynolds number of $Re \approx 1.7 \times 10^4$, which is comparable to our experiments. At the outlet, we have imposed an outflow boundary condition. Flow and pressure conditions at the exit plane are extrapolated from the domain in according to a fully-developed flow assumption -- i.e. zero diffusion flux is assumed in the direction normal to the outflow plane -- and do not affect the upstream flow. We use the Algebraic Wall-Modeled LES (WMLES) SGS model, which is a hybrid RANS/LES method. The eddy viscosity approaches a RANS formulation in areas near walls where turbulence is in equilibrium but poorly resolved, transitioning to an LES formulation in non-equilibrium regions with well-resolved turbulence~\cite{fluenttheoryguide}.
The lateral, bottom, and top planes were then simulated as smooth walls with no-slip boundary conditions.

An adaptive timestep is used such that the Courant number corresponding to the smallest element is $0.5$, with an initial timestep of $1 \times 10^{-6}$ s. Once converged, the timestep is approximately $\Delta t \approx 2 \times 10^{-5}$ s. At $t = 20$ s, the mean and root mean squared (RMS) of the velocity and turbulence kinetic energy (TKE) are sufficiently stabilized, i.e. the change over time for each metric, averaged over the domain at VP2, is negligible, indicating that initial conditions no longer impact the flow. At this point, we begin collecting statistics. We output statistics every $100$ time steps while the simulation runs for an additional $20$ seconds, up to $t = 40$ s. All time-averaged data is from this period, which covers about $10 \times 10^5$ timesteps, and about $1,000$ large eddy turnovers, given a large eddy turnover time of  $\tau_{\text{ed}} \approx H/u_{\infty} = 0.0747$ m $/3.5$ ms$^{-1}$ $\approx 0.02$ s for the main simulations.

\subsection{Kirigami simulation}\label{sect:kirisimmethods}
To simulate flow around kirigami (Sect.~\ref{sect:kirigami}) we first perform a buckling analysis using the commercial package ABAQUS/Explicit 2022. The kirigami geometry is created to match the experiment, with cuts modeled as thin rectangular slits (the cut thickness is $0.25\%$ of the sheet length.) The model is discretized with 3D, quadrilateral shell elements (S4R), which are refined in the vicinity of cuts.  A hyperelastic, isotropic, incompressible material model is used (Neo-Hooke), with the material constant C10 set to 0.46 MPa. Rigid, sacrificial plates are attached with tie constraints to opposite ends of the kirigami sheet, acting as clamps. An encastre boundary condition is applied to one of the attached plates, so that one end of the kirigami sheet remains fixed. A horizontal displacement is applied to the other end, with a small vertical displacement component (magnitude $0.02\%$ of the horizontal displacement component) applied as a bias to cause buckling. Once the buckling mode is identified, we proceed with a dynamic, explicit simulation to obtain the postbuckled shape. 

We export the deformed mesh and, following an intermediate re-meshing step performed using Grasshopper/Rhino, we import the buckled kirigami to Ansys Fluent as a rigid baffle surface. We do not model the clamps used in experiments, but orient the midline of the kirigami surface at the same height as that in the experiment, i.e. $z = 1.25 H$. A fluid mesh is then generated, and the LES is carried out in the same way as for louvers, as described in Sect.~\ref{sect:lesmethods}. 

\section{Validation} \label{sect:validate}
To understand the limitations of these LES and justify the use of simulations for obtaining 3D vector fields and performing a parametric study of louver geometry, we first compare results from experiments and simulations. In Fig.~\ref{fig:validate}a\&b, we show the normalized, time-averaged streamwise ($\overline{u}$) and vertical ($\overline{w}$) velocity contours for $N=3$ louvers from experiments and simulations. All data have been normalized by the mean streamwise inflow velocity $\overline{u}_{\text{in}}$, measured at a distance $0.05 L$ upstream (to the left in all plots) and $0.1 H$ above the box, at $y=0$ (i.e. the middle of the box, width-wise). For all simulations, $\overline{u}_{\text{in}} \in [2.28,2.83]$ m/s, and in experiments, $\overline{u}_{\text{in}} \in [2.35,5.50]$ m/s (see Table 1.) From Fig.~\ref{fig:validate}a\&b, we see that the main flow features observed in experiments are also captured in simulations, with mild deviations that are more pronounced near wall boundaries. The mean inflow conditions also appear similar between experiments and simulations, despite that small screws in the box used in experiments produce locally higher upward velocities at the windward inlet. 

For a closer look at differences between experiments and simulations, we also compare profiles of time-averaged velocity components, and the 2D ($x-z$) root-mean square (RMS) velocity (calculated with respect to the time-averaged velocity), $\sigma$, from experiments and simulations with $N=3$ in Fig.~\ref{fig:validate}c-f. We note that Fluent only provides the resolved parts of the variances, and as such only the resolved parts are used throughout this paper. 
We consider vertical profiles at an upstream location ($0.1 L$ upstream of the box), and five locations within the box, i.e. $x = \{0, \pm 0.07, \pm 0.12 \}$ m (where the interior length of the box $L_{\text{in}}$ spans $0.2858$ m, centered at $x=0$.) In particular, $x0 = -0.15$ m, $x1 = -0.12$ m, $x2 = -0.07$ m, $x3 = -0$ m, $x4 = 0.07$ m, and $x5 = 0.14$ m. We also plot each quantity along horizontal profiles inside the box at $z=H$. This profile is relevant to the computation of the air exchange rate (ACH) in Section ~\ref{sect:ach}. To account for the irregular grid, an unweighted rolling average with a window of $5$ grid points has been applied to LES data in profiles, using the rolling function from the Pandas package in Python. 

In experiments, we observe some vibration of louvers due to wind loading. While this effect would likely need to be damped for real-world designs, the overall agreement between our PIV and LES data in the vicinity of louvers leads us to conclude that the effect of small structural motions on air flow and ventilation is negligible.

We also note that small gaps exist in PIV data where vision was obstructed. The mean velocity profiles, for both $u$ and $w$ agree well in general, particularly inside the box, i.e. at $z/H \leq 1$. A difference in inflow conditions can be seen by the $\sigma$ profile in Fig.~\ref{fig:validate}e, where turbulence is lower in the LES. (This was not significantly improved by imposing turbulent inflow; see Appendix B.) This effect carries through the center of the box, creating larger discrepancies in the RMS velocities compared to the means. However, additional turbulence generated by the interaction with louvers is captured better in the downstream profiles, and the LES and experimental RMS converge in the horizontal profile near the end of the box. This suggests that the RMS velocity discrepancies result largely from a mismatch in inflow turbulence conditions, including internal turbulence generated by wall shear in the wind tunnel. 

Overall, from the profiles in Fig.~\ref{fig:validate}, we conclude that the general trends for the velocity are consistent between experiments and simulations. In addition, we show in Sect.~\ref{sect:ach} that the net air exchange rates between the box and exterior, computed from the LES and PIV measurements, are in good agreement for an open box and one with three louvers. As this metric is our primary concern, and inflow conditions will vary widely in any realistic setting, we conclude that LES is fit to perform a parametric study of louver geometry and orientation. We vary the number of louvers, $N=\{0,2,3,4,5,6,7,14\}$, maintaining that the louver width $\ell = L_{in}/N$, and $\theta = 45 ^\circ$. (We did not simulate $N=1$ louver, as this geometry, i.e. $\ell = L_{in}$, intersects the base of the box when oriented at $\theta = 45 ^\circ$.) We also vary $\theta$ and $H$ for $N=3$. We discuss the results, and implications for ventilation, in the following sections.

\section{Effect of louver geometry on ventilation}\label{sect:results}
\subsection{Flow structure}
\begin{figure*}[t] 
	\centering
	\includegraphics[width=\linewidth]{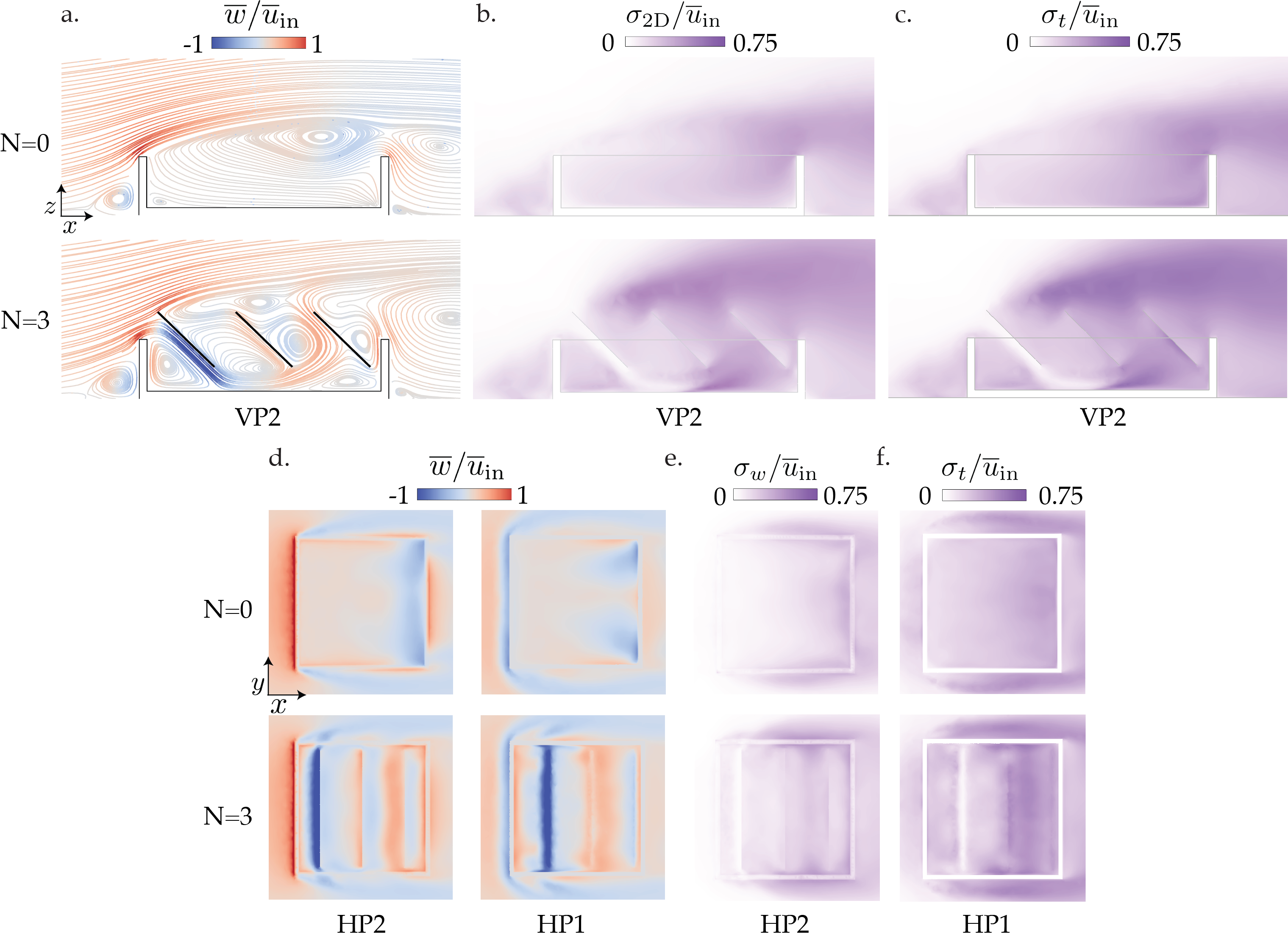}
 	\caption{Comparison of time-averaged velocity data normalized by the mean inflow velocity, from LES for $N=0$ (top row in a.-d.) and $N=3$ louvers (bottom row in a.-d.). The addition of louvers reverses the bulk flow pattern, and draws free air into the box, generating additional turbulence. a. 2D streamlines at VP2 (vertical mid-plane), colored by normalized vertical velocity. b. 2D ($x-z$) RMS velocity pseudocolor plot at VP2. c. Total (3D) RMS velocity (calculated with respect to the time-averaged velocity) pseudocolor plot at VP2. d. Mean vertical velocity pseudocolor plot at HP2 ((horizontal plane at top of box; left) and HP1 (horizontal plane at half of the box height; right). e. Vertical RMS velocity pseudocolor plot at HP2. f. Total RMS velocity at HP1. 
	} 
	\label{fig:allplanes}
\end{figure*} 
In addition to permitting a parametric study, our simulations offer a view of 3D velocity fields that are difficult to access in experiments. To begin to understand the effect of louvers on air flow, in Fig.~\ref{fig:allplanes}, we compare the open box ($N=0$) to that with $N=3$ louvers in three planes: VP2, the vertical midplane, HP2, the horizontal plane at the top of the box ($z=H$), and HP1, the horizontal plane at $z = H/2$. We show the mean velocity streamlines, the 2D (x-z) RMS velocity, and the total (3D) RMS velocity, $\sigma_t$, for VP2, in Fig.~\ref{fig:allplanes} a.-c. For the horizontal planes, we compare the mean vertical velocity in Fig.~\ref{fig:allplanes}c. In addition, we show the vertical RMS velocity $\sigma_w$ at HP2 in Fig.~\ref{fig:allplanes}d (as this contributes to the air exchange through the top of the box), and $\sigma_t$ at HP1 in Fig.~\ref{fig:allplanes}d (as this may be relevant to turbulent mixing at pedestrian or sitting level).

From these plots, it is clear that the relatively large, protruding louvers alter flow patterns significantly. For the open box, a shear layer develops above the box once incoming wind meets the windward face (see Fig.~\ref{fig:allplanes}a. This creates large-scale recirculation, driving air to enter primarily through the back of the box, and exit through the front. This air that exists at the front can then be re-entrained at the back to re-enter the box, reducing effective ventilation. Flow is relatively weak inside the box for $N=0$ louvers. When three louvers are present, air is drawn into the box by the first louver, which intercepts the shear layer. This flow of air, with weak turbulence in our setup, sweeps a portion of the box near the floor, then generates turbulence primarily near the second and third louvers. Air is expelled from the box as it meets the last louver, and between the last louver and the downstream inner wall. The middle louver also directs air upward, though a small amount of this air may reenter the box. Several recirculation regions develop within the box, and some backflow is observed on the windward side of the two downstream louvers, yet air is exchanged in both vertical directions between pairs of louvers.
The reversal of the bulk flow compared to the open box suggests an improved ventilation efficiency; in addition, increased turbulence with $N=3$ louvers could enhance mixing and air exchange inside the box and with the exterior. 

\begin{figure*}[t!] 
	\centering
	\includegraphics[width=0.8\linewidth]{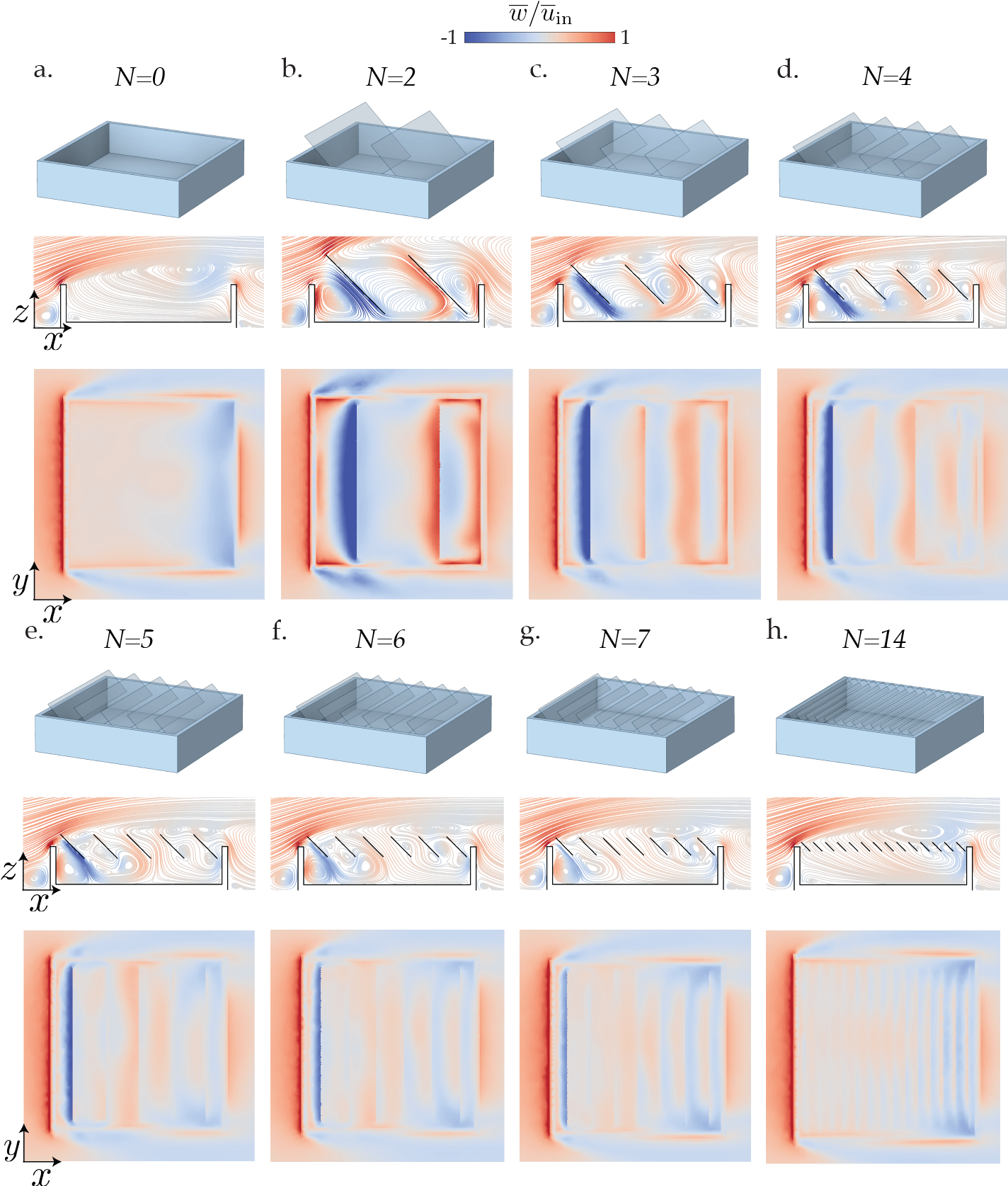}
	\caption{Plots of time-averaged velocity from simulations in center planes, for all louver geometries studied. Larger louvers intercept oncoming flow and draw it into the box, but the effect weakens when the box has more, smaller louvers. a.-h. Top row: Geometry. Middle row: 2D streamlines colored by normalized mean vertical velocity. Bottom row: Pseudocolor plots of normalized mean vertical velocity at HP2 (horizontal plane at top of box).
	} 
	\label{fig:allgeomflows}
\end{figure*} 

With this suggestion that louvers may be able to improve ventilation, we next examine how varying louver geometry impacts flow. In Fig.~\ref{fig:allgeomflows}, we qualitatively compare all louver geometries studied with $\theta = 45 ^\circ$ and $H=7.47$ cm. From the mean streamwise velocity at VP2 and the mean vertical velocity at HP2, we observe that for $N=2$ to $N=7$ louvers ($\ell = 14.29$ cm to $4.08$ cm), we see a similar flow pattern to $N=3$. However, the effect of drawing air into the box weakens considerably as $N$ increases, or $\ell$ decreases. For $N=14$, or $\ell = 2.04$ m -- i.e. with many, small louvers -- the flow essentially recovers the pattern observed for the open box ($N=0$). 

\subsection{Air Exchange Rate (ACH) and efficiency}\label{sect:ach}
\subsubsection{ACH}
To quantify the effect of changing flow patterns with louver geometry on ventilation, we consider the Air Exchange Rate (ACH). The ACH is a common ventilation indicator, representing the volumetric exchange of air with the exterior, per unit time~\cite{Liu2005,LlagunoMunitxa2017,LlagunoMunitxa2018}. In particular, the ACH quantifies the number of box volumes of air flowing out (equal to those flowing in for incompressible flows) per unit time, i.e. the exchange between the interior of the box and the free air aloft. The ACH can be written as: 
\begin{equation}\label{ach}
\text{ACH (s}^{-1}\text{)} =  \frac{1}{H_{\text{in}}} \overbrace{\bigg(\frac{1}{A} \iint_A \underbrace{\bigg( \frac{1}{T} \int_0^T w^+ d t \bigg)}_{\text{temporal averaging}} \: d A\bigg)}^{\text{spatial averaging}} , 
\end{equation}
where $H_{\text{in}}$ is the height of the inner volume of the box; $A$ is the area of the integration plane (in our case, HP2 at the top of the box); $T$ is the total time averaging period, and $w^+$ is the instantaneous positive vertical velocity, defined as:
\begin{equation}\label{wplus}
w^+(\pmb{x},t) = 
    \begin{cases} 
      w(\pmb{x},t) & w(\pmb{x},t) > 0 \\
      0 & w(\pmb{x},t) \leq 0 
   \end{cases}
\end{equation}
for position vector $\pmb{x}$ and time $t$. This expression accounts for turbulent as well as dispersive (mean) exchanges.
Note that $H_{\text{in}} A = V_{\text{in}}$, where $V_{\text{in}}$ is the inner box volume; we have decomposed this term in equation ~\ref{ach} to emphasize the averaging that takes place over the plane. Furthermore, we note that according to this definition, the ACH decreases with increasing box depth, i.e. a deeper structure would have a lower ACH and require a longer time to exchange one volume of air.

\begin{figure*}[t!] 
	\centering
	\includegraphics[width=0.85\linewidth]{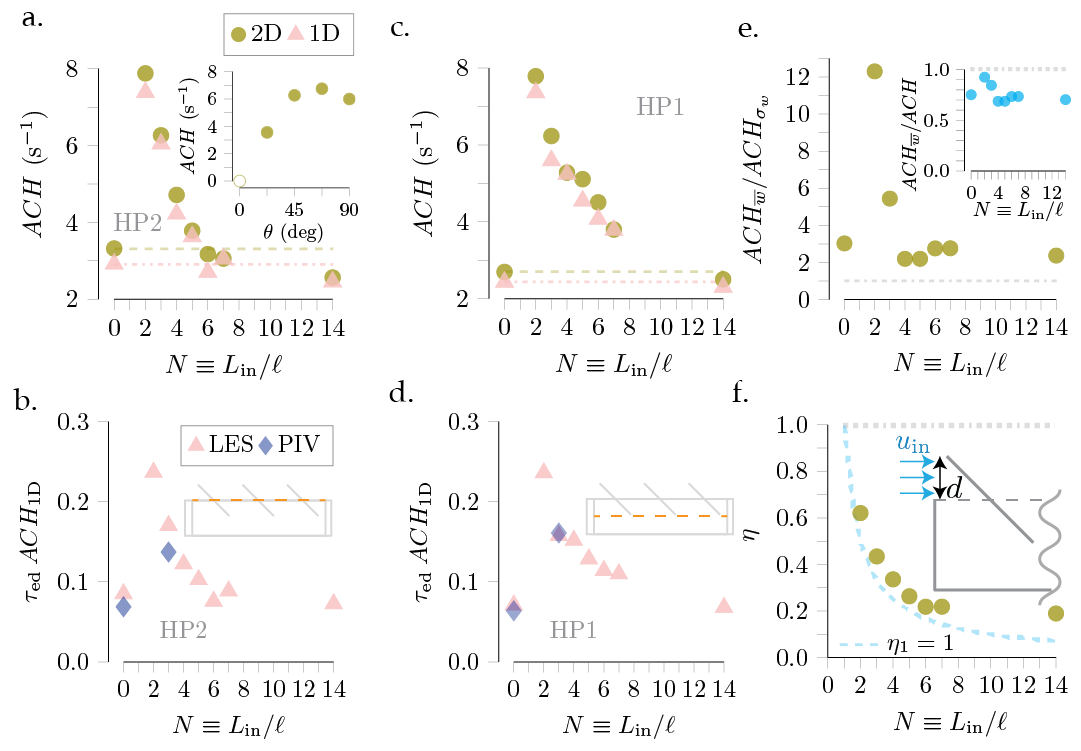}
	\caption{Plots of the 1D and 2D ACH, turbulence contribution, and efficiency measure. For a wide geometric range, the addition of louvers improves the ACH above that for the open box.
    a. \& c. $ACH_{\text{2D}}$ (green circles) and $ACH_{\text{1D}}$ (pink triangles) vs. $N \equiv L_{\text{in}}/\ell$ from simulations. In a., $ACH_{\text{2D}}$ is measured over HP2 (box height), and $ACH_{\text{1D}}$ is taken over the line at the intersection of VP1 and HP2, i.e. $z=H$ and $y=L_{\text{in}}/4$. In c., $ACH_{\text{2D}}$ is at HP1 (half-height), and $ACH_{\text{1D}}$ is at the intersection of VP1 and HP1 ($z=H/2$ and $y=L_{\text{in}}/4$.) 
    a., inset: $ACH_{\text{2D}}$ at HP2 versus the louver opening angle, for $N=3$. (Open marker to guide the eye; no simulation was run for closed box.) 
    b. \& d. $ACH_{\text{1D}}$ normalized by the approximate large eddy turnover time $\tau_{\text{ed}} = H/u_{\text{in}}$, for simulations (pink triangles) and experiments (blue diamonds), at the locations corresponding to a. \& c. (b. top of box; d. mid-height). 
    Insets for b. \& d. indicates the line over which $ACH_{\text{1D}}$ is measured.
    e. Ratio of the dispersive (mean) contribution to the ACH to the turbulent contribution, from the estimates given by \eqref{ACHwbar} and \eqref{ACHsigmaw}. Inset: Fractional contribution of the dispersive fluxes to the total ACH. 
    f. The efficiency $\eta = ACH_V/ACH_V^*$ vs. $N \equiv L_{\text{in}}/\ell$ from simulations. Dotted, gray line: $\eta = 1$. Dashed, blue line: $\eta_1 = ACH_V/ACH_V^1 = 1$, indicating the efficiency if a single louver captures all air according to the approximation in the inset. Inset: Schematic for the calculation of $ACH_V^* = N u_{\text{in}} W d$, the ideal volumetric air exchange rate, and $ACH_V^1 = u_{\text{in}} W d$, the volumetric exchange rate if only a single louver is effective.
    } 
	\label{fig:achplots}
\end{figure*}

Since this calculation is tedious in simulations, the ACH can also be accurately estimated based on a folded normal distribution model of the absolute value of the vertical velocity component $w$ by~\cite{LlagunoMunitxa2018} 
\begin{equation}\label{achfnd}
\text{ACH (s}^{-1}\text{)} = \frac{1}{2 H} \Biggl \langle \sqrt{\frac{2}{\pi}} \sigma_w \exp \Big(- \frac{\overline{w}^2}{2 \sigma_w^2} \Big) -\overline{w} \erf \Big(-\frac{\overline{w}}{\sqrt{2}\sigma_w}\Big) \Biggr \rangle
\end{equation}
where $\sigma_w^2$ is the vertical velocity variance; $\erf$ is the error function; the overbar indicates time-averaging; and angled brackets denote spatial averaging over a horitzontal plane. Eqn.~\eqref{achfnd} is equivalent to~\eqref{ach} if $w$ follows a normal (Gaussian) distribution; this was indeed shown to be a very good approximation for the vertical velocity in Ref. \cite{LlagunoMunitxa2018}, with deviation from Gaussian restricted to the improbable tails of the distribution. As equation~\eqref{achfnd} relies only on the vertical mean velocity and turbulence statistics, it is a less demanding in terms of computational output than equation ~\eqref{ach}.
As such, we use ~\eqref{achfnd} to compute the ACH for our simulations. Considering that the ACH is a bulk parameter and does not guarantee homogeneous ventilation, we consider this metric at both the top of the box, and at mid-box height. Thus, spatial averaging is carried out over the horizontal planes at $z=H$ and $z=H/2$, i.e. HP2 and HP1, for simulations. (Note, when computing the ACH at HP1, the total inner volume is considered, rather than only the volume beneath the midplane.) However, to assess the possibility of computing the ACH in experiments, where it is more practical to measure only along a vertical plane, we also compute \eqref{achfnd} with spatial averaging over the lines at the corresponding heights, and $y=L_{\text{in}}/4$. Thus, whereas we compute $ACH_\text{2D}$ with ~\eqref{achfnd} over a horizontal planar domain, $ACH_\text{1D}$ carries out ~\eqref{achfnd} over a domain restricted to $y=L_{\text{in}}/4$ and $-L_{\text{in}}/2 \leq x \leq L_{\text{in}}/2$, averaging over the points in this linear region. In Fig.~\ref{fig:achplots}a-d, we plot these results for a range of $N \equiv L_{\text{in}}/\ell$. We note that $L_{\text{in}}$ is fixed in our study, so the dimensionless $N$ varies through $\ell$. Due to the relative importance of the upstream louver in capturing wind, which we observe from Fig.~\ref{fig:allgeomflows}, and examine more closely in Sect.~\ref{sect:efficiency}, we surmise that $\ell$ is the crucial parameter for setting the ACH for our box geometry, rather than $N$ itself. 

For both $ACH_\text{2D}$ and $ACH_\text{1D}$, we observe a geometric region wherein the ACH is higher with the addition of louvers than for the open box. In particular, this occurs for $2 \lessapprox N \lessapprox 6$, or $14.29 \lessapprox \ell \lessapprox 4.76$ at HP2, as shown in Fig.~\ref{fig:achplots}a. We note that $N=1$, or larger $\ell$, would likely also increase the ACH, but this is outside of the practical geometric range since opening the louver will cause it to reach the surface, except for very slender and tall boxes. The percent improvement for $ACH_\text{2D}$ ranges from $137 \%$ for $N=2$ louvers to $13 \%$ for $N=5$ louvers. For $N=14$ louvers, the $ACH_\text{2D}$ decreases by $23 \%$. In light of Fig.~\ref{fig:allgeomflows}, these results suggest that the air exchange at the top of the box is increased when louvers reach higher and are thus better able to intercept the shear layer to create strong inflow, and generate flow structures that can sweep the bottom of the box. 


At HP1, we observe a higher ACH for all louvered geometries compared to $N=0$, with the exception of $N=14$. Values are comparable to those at HP2, suggesting that exchange is not limited to the top region of the box. At HP1, the percent improvement in $ACH_\text{2D}$ for louvers, compared to $N=0$, ranges from $189 \%$ for $N=2$ louvers to $41 \%$ for $N=7$ louvers. For $N=14$ louvers, we observe a $7\%$ decrease in $ACH_\text{2D}$. 

The ACH can be further tuned by adjusting the louver angle $\theta$, as shown in Fig.~\ref{fig:achplots}a, inset. For $N=3$ louvers, we observe a non-monotonic dependence when $\theta$ is varied between $22.5 ^\circ$ and $90 ^\circ$ in simulations with respect to the oncoming wind. A peak in $ACH_\text{2D}$ occurs around $\theta = 67.5 ^\circ$. 

Upon comparing $ACH_\text{2D}$ and $ACH_\text{1D}$ in simulations, we find relatively good agreement, as seen in Fig.~\ref{fig:achplots}a\&c. Thus, likely owing to the homogeneous geometry in the $y$ direction within the box, the ACH can be inferred from a vertical plane in experiments or simulations. This also offers an additional means to validate our simulations against the experiments for $N=0$ and $N=3$ louvers. In Fig.~\ref{fig:achplots}b, we plot $ACH_\text{1D}$ normalized by the approximate large eddy turnover time taking the measured inflow velocity into account, i.e. $\tau_{\text{ed}} \approx H/\overline{u}_{\text{in}}$. We observe a higher $ACH_\text{1D}$ in simulations than experiments at HP2, but with a small error (approximately $24 \%$ at HP2 for for $N=0$ and $N=3$ at HP2) 
relative to the difference in the ACH for the different geometries. We note that in experiments, regions of high mean velocity very close to louvers and the top of the box may be visually obstructed due to small oscillations, which could explain the underrepresentation of $ACH_\text{1D}$ in general. At HP1, these obstructions are not present, and experiments and simulations agree well: the error is approximately $10 \%$ for the open box, and $2 \%$ for the $N=3$ geometry.

To understand the relative contributions of mean and turbulent flow to the ACH, we examine the limit of equation \eqref{achfnd} when $\sigma_w$ approaches $0$. This gives an estimate for the dispersive (temporal mean flow,  subsequently averaged in space) exchange as:
\begin{equation}\label{ACHwbar}
    ACH_{\overline{w}} = \frac{1}{2 H_\text{in}} \langle |\overline{w}| \rangle.
\end{equation}

We can then 
simply define the turbulent contribution to the air exchange from equations \eqref{achfnd} and \eqref{ACHwbar} as:
\begin{equation}\label{ACHsigmaw}
    ACH_{\sigma_w} = ACH - ACH_{\overline{w}}
\end{equation}
In Fig.~\ref{fig:achplots}e, we plot the ratio of these estimates of the dispersive and turbulent contributions to the ACH, i.e. $ACH_{\overline{w}}/ACH_{\sigma_w}$, at HP2. In the inset of Fig.~\ref{fig:achplots}e, we also plot the fractional contribution of the mean exchange to the total ACH, i.e. $ACH_{\overline{w}}/ACH$, as computed from equations \eqref{achfnd} and \eqref{ACHwbar}.

From these metrics, we see that while turbulence contributes to air exchange for all geometries, the dispersive contribution to the air exchange is at least double the turbulent contribution for each geometry. This effect is exacerbated when louvers are large: 
for $N=2$ relatively large louvers -- where strong inflow can be observed (see Fig.~\ref{fig:allgeomflows}b) -- over $90 \%$ of the air exchange is due to the mean flow's dispersive effect (approximately 12 times as much as the contribution from turbulence). In contrast, for the open box, and for smaller louvers that do not create strong mean flow chutes, the turbulent contribution makes up about $30 \%$ of the (relatively little) air exchange. Comparing Figs.~\ref{fig:achplots}a \& e, we see that a high ACH correlates with a larger mean flow contribution, enabled by the large louvers. 

Overall, these results confirm that protruding, angled structures can improve air exchange of a space above what it would be in their absence, and more importantly shed light on optimal geometries and angles to maximize this benefit. Another important factor to consider is that the ACH calculation cannot account for air re-entrainment: if a parcel of air leaves the box at the front end and re-enters at the back it is still counted towards ventilation. Given the general circulation patterns observed, the open box will be much more prone to re-entrainment than the boxes with louvers, which thus further motivates the use of louvers for ventilation. 

\subsubsection{Efficiency}\label{sect:efficiency}
From the flow structures in Fig.~\ref{fig:allgeomflows}, it appears that air exchange in our system relies primarily on the ability of the most upstream louver to capture wind. To study the mechanism for ACH increase more closely, we consider an idealized volumetric exchange rate, shown schematically (in 2D) in Fig.~\ref{fig:achplots}f. We reason that if a single louver captures all oncoming air, then the volumetric exchange rate will be approximately $ACH_V^1 = u_{\text{in}} W d$. Here, $d$ is the difference in height between the upper edge of the louver and the top of the box (see Fig.~\ref{fig:achplots}f, inset), i.e. $d = \frac{\ell}{2} \sin (\theta)$, and $W$ is the width of the box as defined in Fig. ~\ref{fig:geom}a. Similarly, if $N$ louvers capture all oncoming air (and, though likely unrealistic, the oncoming velocity is unchanged for downstream louvers), then the ideal volumetric exchange rate will be approximately $ACH_V^* = N u_{\text{in}} W d$. Thus, we define an efficiency parameter $\eta$ as:
\begin{equation}\label{eta}
    \eta = \frac{ACH_V}{ACH_V^*} = \frac{ACH \, V_{\text{in}}}{N\,u_{\text{in}}\,W \,d}. 
\end{equation}
Here, we have defined the observed volumetric air exchange rate $ACH_V$ (with units $m^3/s$) as \eqref{ach}, or \eqref{achfnd} in our computation, multiplied by the inner box volume. Eq. ~\eqref{eta} thus compares the observed ACH to the ideal one, with all louvers participating in capturing freestream air. We also define $\eta_1$, which compares the observed ACH to the single louver case:
\begin{equation}\label{eta1}
    \eta_1 = \frac{ACH_V}{ACH_V^1} = \frac{ACH \, V_{\text{in}}}{u_{\text{in}}\,W\,d}. 
\end{equation}
In Fig.~\ref{fig:achplots}f, we plot $\eta$ versus $N$ for simulations (note that our definition of $\eta$ is meaningless for $N=0$, since $d=0$.) As expected, the efficiency falls well below $1$ for all geometries, and is higher for fewer, larger louvers. The dashed, blue line in Fig.~\ref{fig:achplots}f shows $\eta_1$, i.e. the full efficiency of a single louver. In all geometries, $\eta$ falls above this value. For large $\ell$ (which corresponds to small $N$), $\eta \approx \eta_1$. As louvers become smaller, although $\eta$ decreases overall, the efficiency becomes proportionally higher than $\eta_1$, suggesting that the effect of the single (upstream) louver becomes smaller compared to the overall efficiency. 

Together with the flow structures observed in Fig.~\ref{fig:allgeomflows}, these results indicate that the downstream louvers do not play a first-order role in increasing the ACH. The inaccessibility of the ideal $ACH_V^*$ appears to be primarily due to the upward diversion of the shear layer by the windward box face and upstream louver (as such, the downstream louvers do not experience the horizontal inflow velocity $u_{\text{in}}$.) We speculate that to improve the efficiency, raising the midline of downstream louvers, i.e. changing the global angle of the structure to account for the angle of the shear layer, could promote more air capture downstream. We also note that for larger $L$, i.e. when $L$ exceeds the reattachment length of the shear layer that develops from the first louver, efficiency of downstream louvers may increase. Furthermore, the dominance of the upstream louver suggests that in settings where shade is not required, one could therefore consider omitting downstream louvers. (A single, angled louver recovers a scenario similar to the wind catchers in e.g.~\cite{Chew2017}.) However, realistic environments will face changing wind directions. Thus, a potential improvement for downstream louver(s) lies in adjusting their orientation, which we investigate in the following section.

\begin{figure}[h!] 
	\centering
	\includegraphics[width=0.9\linewidth]{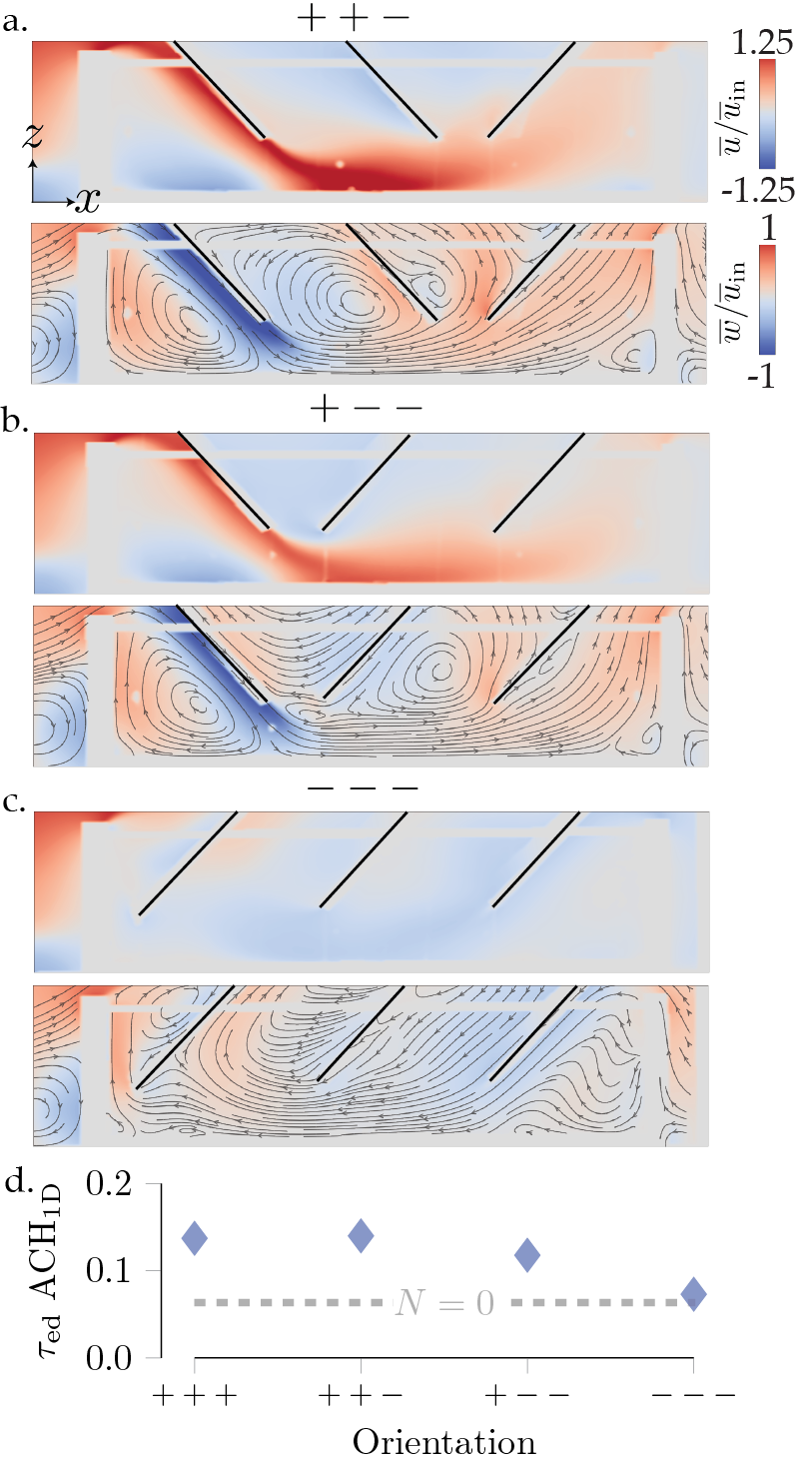}
	\caption{Time-averaged velocity plots and ACH from PIV experiments for $N=3$ louvers with nonhomogenous orientation ($\theta = \pm 45 ^\circ$ for each louver), demonstrating adaptability to changing wind directions. The top rows of a.-c. shows the mean streamwise velocity, and the bottom row shows the mean vertical velocity, with streamlines, for each of three configurations of $N=3$ louvers.
    a. + + - configuration. From left (upwind) to right (downwind), the louver angles are $45 ^\circ$, $45 ^\circ$, and $-45 ^\circ$.
    b. + - - configuration, with louver angles $45 ^\circ$, $-45 ^\circ$, and $-45 ^\circ$.
    c.  - - - configuration. All louvers are angled to $-45 ^\circ$.
    d. Normalized $ACH_{\text{1D}}$ at HP2 versus louver orientation. Dashed, gray line: $\tau_{\text{ed}} ACH_{\text{1D}}$ for the open box.
    The centerlines of louvers have been emphasized with black lines in a.-d.} 
	\label{fig:orient}
\end{figure} 

\section{Mixed louver orientation for changing wind direction}
\label{sect:optimize}
An optimal design for realistic urban environments needs to promote ventilation while accommodating wind approaching from either of two opposing directions, rather than a single prevailing direction. Toward this end, we performed experiments with $N=3$ louvers of nonhomogeneous orientation. In addition to the initial configuration of $45 ^\circ, 45 ^\circ, 45 ^\circ$ ($+++$), we examine $45 ^\circ, 45 ^\circ, -45 ^\circ$ ($++-$); $45 ^\circ, -45 ^\circ, -45 ^\circ$ ($+--$); and $-45 ^\circ, -45 ^\circ, -45 ^\circ$ ($---$). As shown in Fig.~\ref{fig:orient}, the $---$ configuration produces a slightly higher ACH than the open box, but significantly lower air exchange compared to $+++$. In other words, the $+++$ design would not perform well if wind reversed direction. However, both the $++-$ and $+--$ configurations exhibit higher $ACH_\text{1D}$ than $---$, and the $ACH_\text{1D}$ for $++-$ even exceeds that for the $+++$ configuration. 

Comparing Fig.~\ref{fig:orient}a\&b with Fig.~\ref{fig:allgeomflows}, once can see that this improvement is likely due to the ability of the downstream louvers to guide flow out of the back of the box, rather than blocking it. As the $++-$ and $+--$ configurations are the reverse of one another, this design would allow for strong air exchange for opposing wind directions. We speculate that like the $---$ configuration, a $90 ^\circ$ wind would likely provide only small improvement compared to the open box for ventilation. However, we did not study a wider range of wind angles, for which ventilation likely would not be as efficient. One could consider tiling louvers of differing orientations to improve ventilation from multiple wind angles. Furthermore, enhanced wind steering could likely be achieved with nonhomogeneous angular magnitudes, e.g. $67.5 ^\circ, 45 ^\circ, -30 ^\circ$.

\section{Effect of box height}
\label{sect:height}
\begin{figure*}[t!] 
	\centering
	\includegraphics[width=0.95\linewidth]{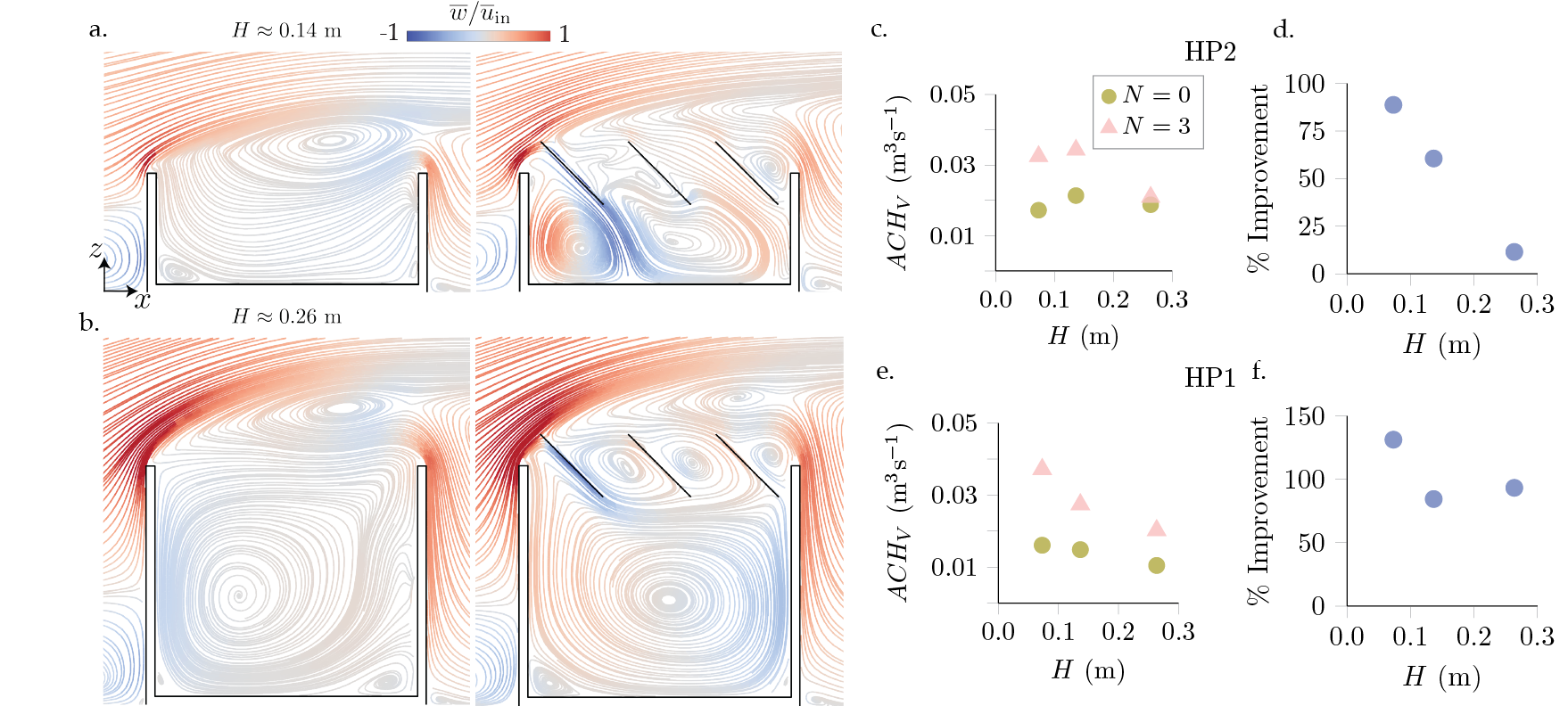}
	\caption{Effect of box height on flow and ACH. a. \& b. Mean streamwise velocity for $N=0$ (left) and $N=3$ (right) louvers for the a. mid-height and b. tallest box geometries studied. c.\& e. $ACH_V$ versus height for $N=0$ (green circles) and $N=3$ louvers at $\theta=45 ^\circ$ (pink triangles) at HP2 (c.) and HP1 (e.) d.\&f. Percent improvement to $ACH_V$ for $N=3$ compared to $N=0$ versus height at HP2 (d.) and HP1 (f.)
 }
	\label{fig:height}
\end{figure*} 
Thus far, we have investigated the basic flow physics for ventilation of a shallow, isolated box with varied geometry of protruding angled features. This has shown that compared to an open box, louvers improve the ACH when the louver size is approximately $0.7 H$ or larger. However, we do not expect that louvers necessarily need to be enlarged proportionally to an increasing box height in order to maintain ventilation improvement. Rather, in realistic urban settings, the geometry of angled features needed for ventilation will likely also depend on factors impacting the location of flow separation and the height of freestream air upstream, e.g. the geometric relationship to surrounding buildings. 

Still, excavating stagnant air from a larger cavity poses a greater challenge, which we investigate next by comparing results for $N=0$ and $N=3$ louvers, with walls of two and four times the initial height (and base thickness constant), i.e. $H = 13.65$ cm and $H = 26.35$ cm, or $H=0.45L$ and $H=0.86L$ ($H=0.25L$ for the initial geometry.) Contour plots of the mean streamwise velocity for these geometries, with corresponding ACH results, are shown in Fig.~\ref{fig:height}. Here, to compare ventilation of spaces of different volumes, we use the volumetric exchange rate $ACH_{V}$. 
In Fig.~\ref{fig:height}c\&d, we also show the percent improvement in $ACH_{V}$ for $N=3$ louvers compared to $N=0$ for each height, at HP1 and HP2. 

From Fig.~\ref{fig:height}, we see that the shear layer for $N=0$ reaches higher above the box, and occurs at a larger angle, as the box height increases. 
However, the upstream louver still captures some air in both taller geometries. Compared to the initial geometry, i.e. $H=0.074$m, the inflow and updraft appear weaker. However, inflow does reach the floor of the mid-height geometry, and the $ACH_{V}$ is comparable to the initial configuration at HP2. This is not the case for the tallest box, and the $ACH_{V}$ is lower for this geometry, suggesting that the flow deviation does not reach the ground, and thus does not get deflected downstream to aid in the ventilating circulation. 
At HP2, the improvement for $N=3$ louvers compared to $N=0$ decreases monotonically with increasing height (Fig.~\ref{fig:height}d), but remains above zero in the range of heights we study.

At HP1, which could be more relevant to inhabitants, $ACH_{V}$ is reduced for both taller heights at HP1, i.e. ventilation is poorer at lower depths as height increases, for both $N=3$ and $N=0$. However, the percent improvement due to louvers is greater than $84 \%$ in all cases at HP1. Considering that the volumes of these taller boxes are larger, and yet their $ACH_V$ values are comparable or lower, their ACH values will be severely lower than the shorter box, indicating poorer ventilation.


\begin{figure*}[t] 
	\centering
	\includegraphics[width=\linewidth]{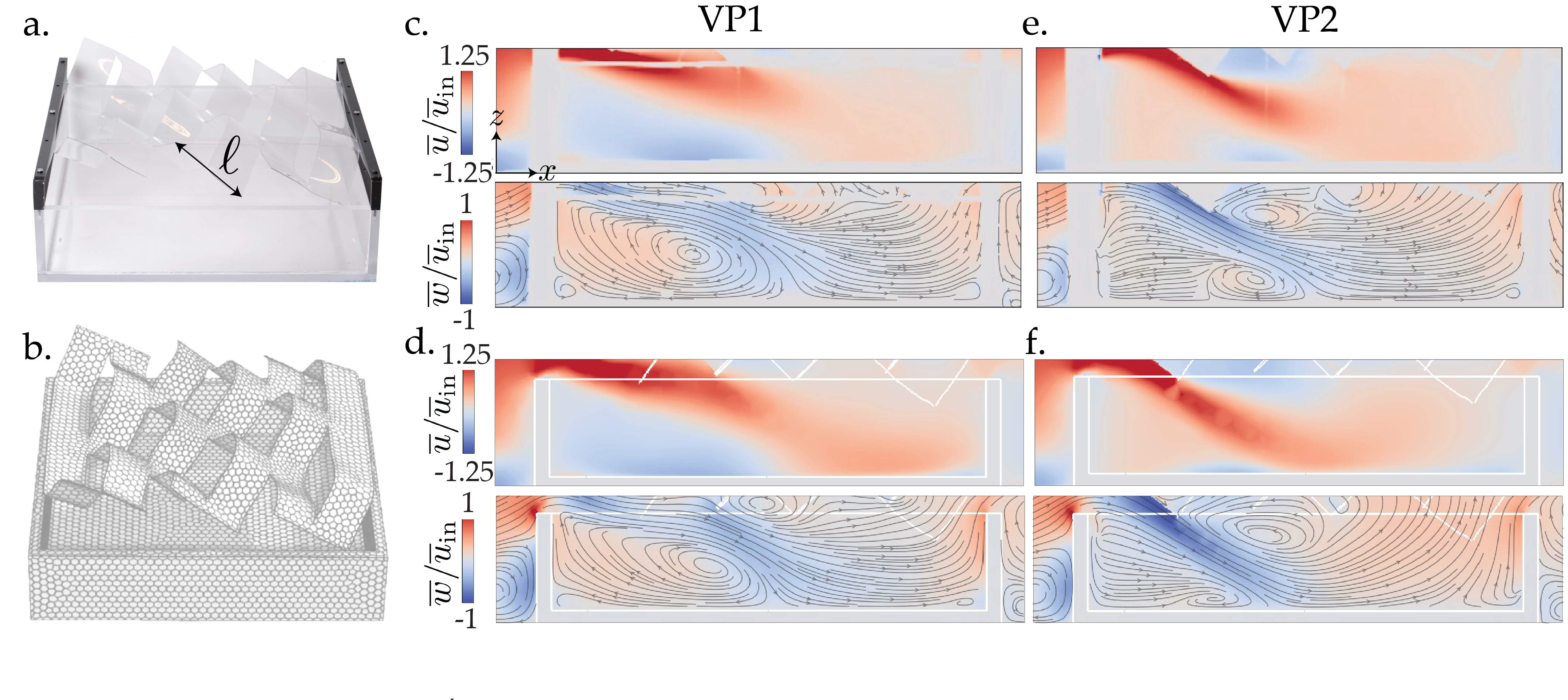}
	\caption{Demonstration of kirigami for ventilation. a. The experimental setup with $20 \%$ strain ($L_y = 25.4 cm$). 
    b. Simulation with surface mesh shown. 
    c.-f. Pseudocolor plots of mean streamwise (top) and vertical (bottom) velocity, with streamlines shown in the bottom plot. 
    c. Experiments, at VP1 (quarter plane). 
    d. Simulations, at VP1 (quarter plane). 
    e. Experiments, at VP2 (midplane). 
    f. Simulations, at VP2 (midplane). 
	}
	\label{fig:kiri}
\end{figure*}

\section{Kirigami demonstration}
\label{sect:kirigami}
Kirigami has a much more complex geometry than louvers, but offers significant advantages, including single-degree-of-freedom actuation via uniaxial stretching; easy, zero-waste construction (linear, parallel cuts); and increased strength and flexibility due to curvature. Kirigami is also widely customizable, and can be bistable~\cite{Yang2018} to accommodate changing wind directions.

In this study, we investigate one simple design, via experiment and simulation, shown in Fig.~\ref{fig:kiri}a\&b. We note that a real-world design should allow for a range of applied stretch, and thus a range of deformation. However, for simplicity, the amount of stretch is fixed by setting the overall length of the sheet to $25.4$cm. Cut sheets are stretched to fit the length of the box, $L$, so the strain (i.e. the amount of stretch compared to the unstretched length of the sheet) is $20$ percent. The cut length is fixed at $12.7$ cm, with two repeated units width-wise and a horizontal spacing of $2.54$ cm. Informed by the results for louvers, we create three repeated units length-wise, so $N=3$. However, $\ell = 7.82$ cm, which is more comparable to the geometry of $N=4$ louvers. Because clamped boundary conditions are applied to the ends of the kirigami sheet, sections of the sheet near the clamps are prevented from buckling upward. Thus, to allow airflow in these regions, we also raise the kirigami sheet by $0.25H$. Thus, the length $d$ is $0.86 H$ measured at the upstream unit cell. 

\begin{figure}[h!] 
	\centering
	\includegraphics[width=0.8\linewidth]{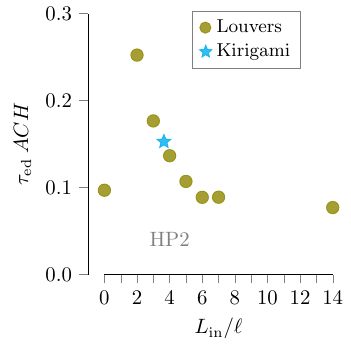}
	\caption{ACH for kirigami compared to louvers. Plot of normalized $ACH_\text{2D}$ at HP2, for louver simulations with $H=7.47$ cm and $\theta = 45^{\circ}$ (i.e. the same data plotted in Fig.~\ref{fig:achplots} a, but normalized by $\tau_{\text{ed}} = H/u_{\text{in}}$; green circles), and kirigami (blue star). Kirigami promotes air exchange in a comparable way to louvers (data at HP2 from simulation.)  
	}
	\label{fig:kiriach}
\end{figure} 

In Fig.~\ref{fig:kiri}c\&d, we show the mean streamwise and vertical velocity in VP1 from the experiment and simulations. Unlike louvers, the kirigami geometry is not homogeneous in the $y$ direction. Thus, we also examine VP2 in Fig.~\ref{fig:kiri}e\&f. In both planes, experiments and simulations agree sufficiently well, despite that the clamps used in experiments are not modeled in experiments. In the vertical midplane (VP2), the louver geometry approximates the kirigami section well, and indeed we observe similar flow patterns compared to e.g. Fig.~\ref{fig:validate}a\&b. Where flow meets angled chutes, we observe inflow, which we have seen increases the ACH, and in general, exiting air finds spaces between kirigami unit cells. We note that due to boundary effects -- which are dominant since the number of repeated units ($3$) is small -- we note that the equilibrium angles for each unit are nonhomogeneous~\cite{Taniyama2019}. At the free edge, the angles are approximately $28$, $41$, and $30 ^{\circ}$, from upstream to downstream. This effect could perhaps be improved by altering the cut pattern near the boundaries~\cite{Taniyama2021}. Even with this suboptimal aspect of the design, in Fig.~\ref{fig:kiriach}, we see that the ACH is significantly above that for the open box, and is very comparable to louvers. This result affirms the use of the simplified, louver geometry for guiding kirigami design, and suggests that kirigami can offer a viable solution for increasing ventilation. Overall, these findings point to a strong potential for kirigami to be further optimized for ventilation, which we pursue in forthcoming work~\cite{Stein-Montalvoinprep}.

\section{Conclusions and future work}
\label{sect:conclusions}
In this work, we have demonstrated that retrofitting spaces with protruding, angled, features offers a route to wind steering for improved ventilation. 
As such designs can also reasonably be expected to provide shade~\cite{Horner2014}, this offers a route to simultaneously and adaptively improving thermal comfort and ventilation in urban spaces. 
Specifically, we performed a parametric LES study, supported by wind tunnel experiments with PIV measurements, of an idealized, shallow container, to identify a geometric regime where louvers can increase the ACH above what it would be for an equivalent box with an open top. Additionally, we demonstrated that for our system, the ACH across a horizontal plane can be well-approximated by data over a line at the same height. For similar, quasi-2D flows, this 1D ACH measure can be used to glean data from 2D planes in experiments, and simplify computational postprocessing. We also considered the relative contributions of mean and turbulent air exchange. We find that dispersive fluxes dominate over turbulent exchanges in general, and strong mean inflow correlates with a higher air exchange rate. 
Thus, the key mechanisms for improving ventilation are: louvers (1) intercept free stream air, drawing it into the box, (2) direct it along a surface, to be expelled downstream, and (3) create turbulence.
By doing so, the main direction of flow inside the box with louvers remains the same as the incoming wind. This contrasts with the bulk recirculation that occurs in an open box that may promote re-entrainment. Louvers, therefore, increase air exchange rates and reduce recirculation.  

To quantify the effectiveness of air exchange based on the capture of horizontal wind, we also define an efficiency measure, $\eta$. The values of $\eta$ for the geometries we study, along with flow patterns, suggest that the upstream louver plays a dominant role in increasing the ACH, as oncoming air deflects upward, out of reach of downstream louvers. Efficiency could perhaps be increased by introducing a negative global angle to the louvered or kirigami structure, so that downstream louvers reach higher than the upstream one, and thus can intercept the shear layer. However, the apparent ineffectiveness of the downstream louvers in this system also free them to be tailored for accommodating wind from different directions. Thus, with an understanding of the mechanisms for air exchange, we then demonstrated with PIV experiments a route to optimizing ventilation for multi-directional wind by mixing louver orientation. 

We also showed with LES that even for taller box geometries, air exchange can be increased with louvers, though results suggest that ventilation at pedestrian level would likely need further intervention for deeper canyons. Finally, we performed experiments and simulations using kirigami, and found strong potential for louver-like ventilation, evidenced by inflow at chutes and an increase in the air exchange on par with that produced by louvers. 

By focusing on ventilation of an isolated box, we have been able to uncover the basic flow physics, which may be a reasonable approximation for a standalone streetery within an urban canyon, or perhaps an interior courtyard. Besides our limited study of the effect of a larger $H/L$ ratio on ventilation, we did not investigate different structural morphologies or aspect ratios, which could influence flow deflection~\cite{Paterson1990,Kono2016,LlagunoMunitxa2017,LlagunoMunitxa2018,Wang2020}, and thus the effectiveness of ventilation. While our findings confirm the potential of kirigami-inspired ventilation of interior and exterior spaces, generalization of the quantitative aspects to other urban topographies requires characterizing the exterior flow field, which is determined by upstream features such as buildings and other obstacles~\cite{Buccolieri2022}, and canyon geometry~\cite{Oke1988}. We did not attempt analysis in such realistic settings in this work. However, we expect that the wind-steering mechanism could be accessed in other urban archetypes fitted with angled features. It has been shown that the presence of a single, upwind, angled windcatcher can increase airflow in idealized, 2D canyons with an aspect ratio of one~\cite{Chew2017}, and a similar design can significantly reduce pollutant concentration in a realistic urban street~\cite{Lauriks2021}. Thus, in geometries that do not produce as much flow separation as an isolated building, louver-generated ventilation could perhaps be even more accessible. For example, a relatively shallow, cavity-like canyon geometry that undergoes skimming flow would likely benefit from the addition of angled louvers. An understanding of how building or canyon geometry affects the reattachment length of shear separation could also guide design optimization. This could include altering the global height or angle of the midline of louvers, for example, which we kept constant at the top of the box. 


Additionally, the Reynolds number in our experiments and simulations is approximately one order of magnitude below what would be anticiapted in typical urban settings. We do not expect to see a major change to the mean flow features we have focused on at higher $Re$, but imposing plausible turbulence characteristic of the atmospheric boundary layer (ABL) would likely be useful. Furthermore, our study focused on the ACH, which is a bulk ventilation quantity. A more detailed analysis of heterogeneity of air exchange and interior flow structures -- including recirculation regions that can develop and result in re-entrainment -- e.g. beneath the upstream louver, in our geometry -- would be beneficial for design. 

Finally, we did not investigate mechanical factors that could pose significant challenges to real-world feasibility, e.g. the loading response of louvers or kirigami (which we treated as rigid in the present study) to lift forces from strong wind or gusts, structural supports, or actuation options. 
From a fluid mechanics perspective, kirigami poses unique challenges due to its nonhomogenous 3D geometry. However, the mechanisms for wind-steering identified in the present work will inform these kirigami designs, which we will develop further in forthcoming work~\cite{Stein-Montalvoinprep}, and could also guide the development of many other structures aimed at improving urban ventilation. 

\section*{Acknowledgments}
The authors are grateful to Lingxiao Yuan for help with finite element simulations of kirigami, to Luc Deike and Claudia Brunner for helpful discussions, to Cam My Nguyen for help with architectural drawings, to Joseph Vocaturo and Larry McIntyre for help with the experimental setup, and to Michael Vocaturo for access to the wind tunnel.
LSM acknowledges financial support from the Princeton Presidential Postdoctoral Research Fellowship and the Momental Foundation. LSM \& EBZ acknowledge support from the Army Research Office under contract W911NF-20-1-0216. The simulations presented in this article were performed on computational resources managed and supported by Princeton Research Computing, a consortium of groups including the Princeton Institute for Computational Science and Engineering (PICSciE) and the Office of Information Technology's High Performance Computing Center and Visualization Laboratory at Princeton University.

\section*{CRediT authorship contribution statement}
\textbf{LSM}: Conceptualization, Investigation, Software, Writing - Original Draft. \textbf{LD}: Investigation, Writing - Review \& Editing. \textbf{MH}: Supervision, Writing - Review \& Editing. \textbf{SA}: Supervision, Writing - Review \& Editing. \textbf{EBZ}: Conceptualization, Supervision, Writing - Review \& Editing.



  \bibliographystyle{elsarticle-num} 
  \bibliography{louverbib}
  
  \clearpage

\onecolumn 
\section*{Appendix A: Wind tunnel velocity profiles} \label{sect:SIwt}
\begin{figure}[h!] 
	\centering
		\includegraphics[width=0.9\linewidth]{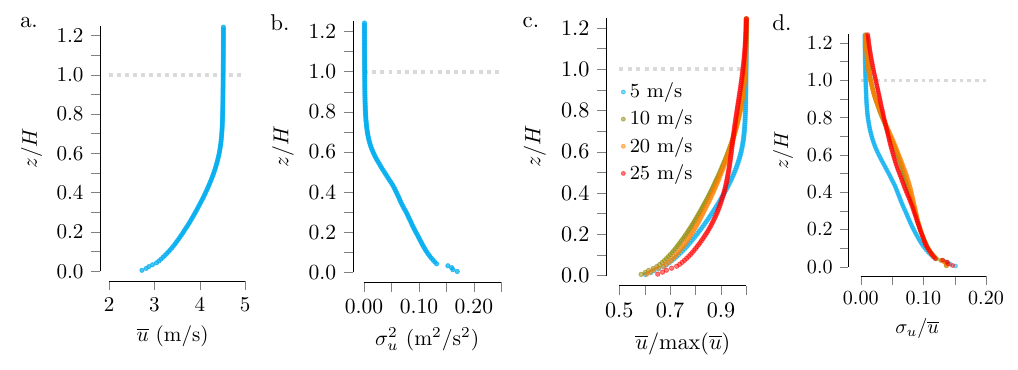}
	\caption{Velocity profile measured in the empty wind tunnel, averaged over location 1 (i.e. the location of the upstream portion of the box during experiments; see Fig.~\ref{fig:geom} in main text.)
    a. Streamwise velocity versus the vertical coordinate, normalized by the box height, for the fan pitch setting used in experiments. 
    b. Streamwise velocity variance for the same experiment as a.
    c. Streamwise velocity normalized by the maximum of $\overline{u}$ in the frame. Legend indicates max($\overline{u}$). 
    d. RMS streamwise velocity normalized by the mean streamwise velocity for the same experiments as in c. 
	} 
	\label{fig:wtbl}
\end{figure} 
To characterize the air flow in our experiments, we took additional measurements at location 1 in the wind tunnel (see Fig.~\ref{fig:geom} in the main text), i.e. the location of the upstream portion of the box during experiments. Velocity is controlled by setting the wind tunnel fan pitch, and in Fig.~\ref{fig:wtbl}a-b, we show profiles for the fan pitch setting used in the main experiments, which result in a freestream velocity of approximately $5$ m/s. The mean streamwise velocity is plotted in Fig.~\ref{fig:wtbl}a, and the velocity variance is shown in Fig.~\ref{fig:wtbl}b. 

To check the sensitivity of the profiles to different velocities, in Fig.~\ref{fig:wtbl}c we also  plot the mean streamwise velocity normalized by its maximum value within the plotted frame (i.e. above the boundary layer, at $z/H \approx 1.25$), for four different fan pitch settings. This results in freestream velocities of approximately $5$ m/s (blue curves, as used in main experiments), $10$ m/s, $20$ m/s, and $30$ m/s. In Fig.~\ref{fig:wtbl}d, we plot the the normalized RMS velocity, i.e. $\sigma_u/\overline{u}$, for the same experiments. 

We observe relatively good collapse of the normalized mean flow profiles, and of the normalized turbulence profiles for the three higher velocities. The profiles appear to be smoother at higher velocities. We were constrained to remain in this lower velocity range due to the fragility of louvers. (We note that the kirigami sheets we tested can withstand higher forces.) 

\section*{Appendix B: LES grid \& inlet condition sensitivity} \label{sect:SIgrid}
\begin{figure}[h!] 
	\centering
	\includegraphics[width=0.8\linewidth]{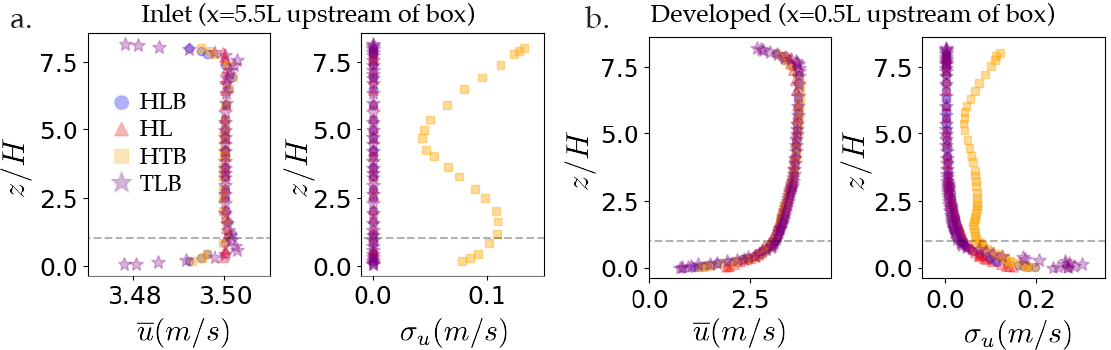}
	\caption{LES sensitivity study for $N=3$ louvers: inflow conditions.
		a. Mean streamwise velocity profiles (left) and RMS streamwise velocity (right) at the inlet for $N=3$ louvers, for LES with varied conditions according to Table~\ref{table:sensitivity}.
        b. Mean streamwise velocity profiles (left) and RMS streamwise velocity (right) at half of the box length upstream of the box, where the boundary layer has developed, for $N=3$ louvers for the same simulations as in a.  
	} 
	\label{fig:sensitivityfull}
\end{figure} 

\begin{figure*}[t] 
	\centering
	\includegraphics[width=0.8\linewidth]{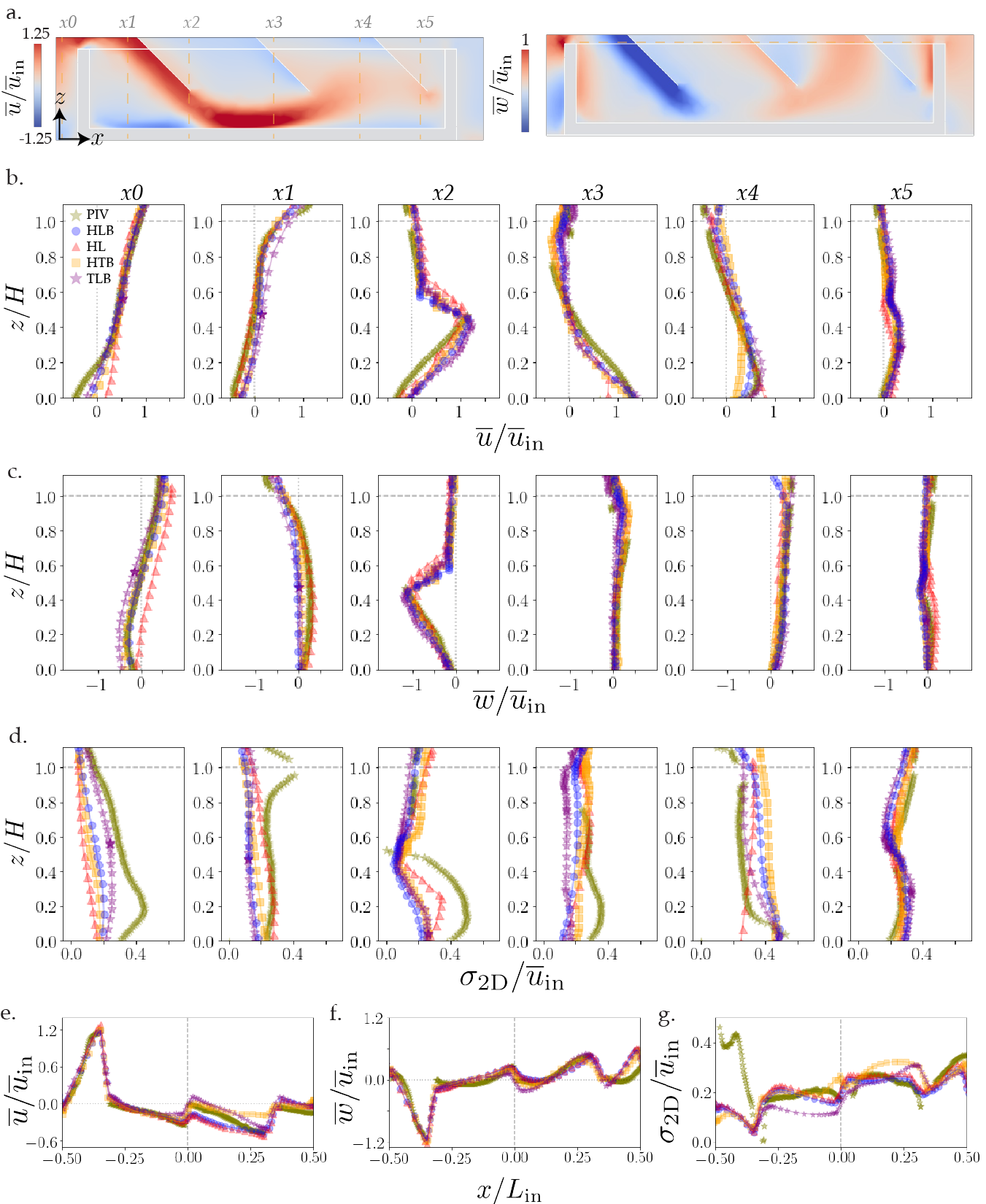}
	\caption{LES sensitivity study for $N=3$ louvers, in the vicinity of the box.
		a. Pseudocolor plots at VP1 of time-averaged velocity data for the HLB simulation used in the main text, with $N=3$ louvers. Vertical profile lines $x0-x5$ are delineated in the streamwise velocity plot on the left, and horizontal profile line is delineated in the vertical velocity plot on the right. 
        b.-e. Mean velocity profiles for $N=3$ louvers, comparing PIV (green stars) to LES with varied conditions according to Table~\ref{table:sensitivity}.
		 Vertical profiles in b.-d. are taken at $x0 = -0.15$, $x1 = -0.12$, $x2 = -0.07$, $x3 = -0$, $x4 = 0.07$, and $x5 = 0.14$ m. Horizontal profiles in f. are at $z=H$, i.e. at the top of the box.
		 b \& e. Streamwise velocity. 
		 c \& f. Vertical velocity.
		 d. \& g. Total 2D ($x-z$) RMS velocity. 
	} 
	\label{fig:sensitivity}
\end{figure*}

A LES sensitivity analysis was performed for the box with $N=3$ louvers, wherein we studied the effects of the grid resolution, boundary layer refinement, and inlet condition on the time-averaged mean and RMS velocities (defined with respect to the mean), as well as the air exchange rate. The configurations are summarized in Table 2, and the performance of the different meshes are shown in Figs.~\ref{fig:sensitivityfull} and~\ref{fig:sensitivity}.

In Fig.~\ref{fig:sensitivityfull}, we show the mean streamwise and RMS streamwise velocity profiles just beyond the inlet, and half of the box length upstream of the box, for each of the conditions summarized in Table 2. We note that the data, exported from Ansys CFD-Post, uses conservative, i.e. control volume values (rather than hybrid values, which represent the specified boundary condition values, i.e. zero velocity at the walls.) 

\begin{table*}[h!]
	\label{table:sensitivity}
	\centering
	\begin{tabular}{cccccccccc}
		\toprule
		& \multicolumn{1}{c}{Elem. shape} & \multicolumn{1}{c}{Inflow} & \multicolumn{1}{c}{BL} & No. cells & \multicolumn{1}{c}{Comp. cores} & 
  \multicolumn{1}{c}{GB/core} & 
  \multicolumn{1}{c}{Run time (days)} & 
  \multicolumn{1}{c}{$ACH_{\text{2D}}$ ($s^{-1}$)}& \multicolumn{1}{c}{$ACH_{\text{1D}}$ ($s^{-1}$)} 
  \\
		\midrule
		HLB & Hex. & Lam. & Res. & $9.7 \times 10^4$ & $4$ & $1$ & $4$ & 6.27 & 6.04\\
		HL & Hex. & Lam. & Unres. & $6 \times 10^4$ & $4$ & $1$ & $1$ & 6.06 & 6.45\\
		HTB & Hex. & Turb. & Res. & $6 \times 10^4$ & $4$ & $1$ & $1$ & 6.45 & 6.63\\
		TLB & Tet. & Lam. & Res. & $4 \times 10^5$ & $10$ & $4$ & $13$ & 5.78 & 6.35\\
		\bottomrule
	\end{tabular} 
	\caption{Mesh and inflow configurations studied for $N=3$ louvers in sensitivity analysis. The columns refer to the surface element shape (``Hex." for hexagonal or ``Tet." for tetrahedral), inflow condition (``Lam." for laminar or ``Turb." for turbulent), boundary layer meshing (``Res." for resolved or ``Unres." for unresolved), number of cells, number of computational cores used in parallelization, computational memory in GB per core, the time for the simulation to run for $40$s, the 2D ACH at HP2, and the 1D ACH at $z=H$ and $y=0$.}
\end{table*}

For the parametric study in the main text, we selected a hexagonal surface mesh (polyhedral volume mesh) in Fluent, with laminar inflow and a resolved boundary layer, which we refer to as HLB, for ``hexagonal, laminar, boundary layer." The grid is refined locally by setting the element face size to $0.007$ on the louvers and box with a growth rate of $1.2$, whereas the far-field maximum element size is $6.5 \times 10^{-3}$ m$^2$. This corresponds to approximately 12 grid cells along the height of the box, 16 for each of $N=3$ louvers, and about $9.7 \times 10^4$ cells in total. For the parametric study in the range of $N=0$ to $N=14$ louvers, the number of grid cells ranged from approximately $7.7 \times 10^4$ to $9.8 \times 10^4$. 

We found that adjusting the local face size minimally affected results; these results are not contained herein. We also did not find significantly improved agreement with experiments by adding turbulence in the inflow (``hexagonal, turbulent, boundary layer"; HTB). Turbulence was imposed using the Vortex Method with $190$ vortices generated at each time step, with their intensity and size determined by the imposed local value of the turbulent intensity (set to $5 \%$), and the molecular to turbulent viscosity ratio (set to $1000$). As can be seen from Fig.~\ref{fig:sensitivityfull}, the turbulence from the inlet (yellow squares) becomes less dominant downstream, and internally generated turbulence is instead more relevant in the vicinity of the box. However, the choice of whether to resolve the boundary layer was more consequential, as seen by comparing HLB to the HL (``hexagonal, laminar") condition near solid boundaries in Fig.~\ref{fig:sensitivity}.

Choosing a tetrahedral mesh (``tetrahedral, laminar, boundary layer"; TLB) provided somewhat improved results compared to the selected HLB, but at a much larger computational cost for the same local sizing. While TLB may be preferred in some cases, improvements are minimal, so the trade-off favors the HLB configuration for our purposes. Furthermore, we note that the values for the 2D ACH, which is the central metric for our study, have a range of $0.67$ (1/s). This error range is much smaller than the difference between $ACH_{\text{2D}}$ for $N=3$ louvers compared to the open box, which is approximately $3$ (1/s).




\end{document}